\documentclass[11pt,a4paper]{article}
\usepackage{lmodern}
\pdfoutput=1
\usepackage{jheppub}
\usepackage{rotating}
\usepackage{array}
\usepackage{amsmath}
\usepackage[normalem]{ulem}
\usepackage{slashed}
\usepackage{booktabs}
\usepackage[pdftex,table,usenames,table]{xcolor}
\usepackage{enumerate}
\usepackage{parskip} 
\usepackage{upgreek}
\usepackage{layouts}
\usepackage{appendix}
\usepackage{siunitx}
\usepackage{paralist}
\usepackage{multirow}
\usepackage{xspace}
\usepackage{tikz}
\usepackage{hyperref}
\usepackage{mathrsfs}
\usepackage{pgfplots}
\usepackage{fontawesome5}
\usepackage{silence}
\usepackage{footmisc}
\usepackage[most]{tcolorbox}
\pgfplotsset{compat=1.15}

\arxivnumber{2601.17119}
\preprint{MPP-2026-4}

\title{NO LESS: Novel Opportunities for Light Exotic Searches at the SPS}

\author[1]{Babette D\"{o}brich}
\author[1]{Jan Jerhot}
\author[2]{Karim Massri}
\author[1,3]{Jonathan L. Schubert}
\author[4]{Tommaso Spadaro}

\affiliation[1]{Max-Planck-Institut für Physik (Werner-Heisenberg-Institut), Boltzmannstr. 8, 85748 Garching bei München, Germany}
\affiliation[2]{Physics Department, Lancaster University, Bailrigg, Lancaster LA1 4YB, UK}
\affiliation[3]{Technical University of Munich, TUM School of Natural Sciences, Physics Department, Chair for Data Science in Physics, 85748 Garching, Germany}
\affiliation[4]{Laboratori Nazionali di Frascati dell’INFN, Via E. Fermi, 54, 00044 Frascati Italy}

\vspace{0.2cm}

\emailAdd{babette@mpp.mpg.de}
\emailAdd{jan.jerhot@cern.ch}
\emailAdd{karim.massri@cern.ch}
\emailAdd{jonathan.schubert@cern.ch}
\emailAdd{tommaso.spadaro@cern.ch}

\abstract{
A powerful way to test models with feebly interacting particles in the MeV to GeV mass range is through proton beam-dump experiments. In this paper, we compare the current sensitivity of CERN’s NA62 experiment running in beam-dump mode with that of a hypothetical experiment using the same detectors in a future CERN ECN3 beam-dump facility. When optimising such an experiment, the geometric setup is particularly relevant for the specific new-physics scenario under study, since different production mechanisms can generate different angular distributions of new particles. We show that even the most minimalistic reconfiguration of the existing NA62 experiment’s detectors can already provide a very competitive sensitivity and collect data immediately after the beam is available.
}


\begin{document}
\maketitle

\section{Introduction}

Testing new physics at moderate (MeV–GeV) mass scales and very weak couplings has gained traction over the past decade: the number of experiments active in this field now numbers in the dozens, and a recent effort to summarise the current status of the field has been made as input to the European Strategy for Particle Physics~\cite{deBlas:2025gyz}. The motivation behind these searches is that a set of portals with different quantum numbers could connect the Standard Model~(SM) to another particle sector, which might, for example, contain a candidate for dark matter. To classify these new-physics scenarios, also referred to as FIPs (feebly interacting particles), a set of “benchmark cases” (\textit{BC}s) has been introduced~\cite{Beacham:2019nyx}. These are now widely established in the literature.

To probe coupling ranges in which such particles are very weakly interacting, and thus long-lived, a detector displaced from the production point is required. An example of a dedicated experiment exploiting proton collisions at the highest energies is FASER~\cite{FASER:2022hcn}, which aims to detect new long-lived particles produced at the LHC. With lower energy but much higher effective luminosity, proton beam-dump experiments are a powerful tool for probing the existence of the above-mentioned portals: they allow higher production rates than collider-based searches and access to a broad spectrum of production processes. Consequently, a number of constraints on portals have been explored by experiments such as CHARM~\cite{CHARM:1985nku,CHARM:1985anb}, NuCal~\cite{Blumlein:2013cua}, and others~\cite{NuTeV:1999kej,Barouki:2022bkt}.

At the time of writing, there are two active experiments capable of probing \textit{BC}s in a proton-dump configuration: DarkQuest at Fermilab~\cite{Apyan:2022tsd,darkquestdata}, which uses a 120~GeV proton beam, and NA62~\cite{NA62:2017rwk} at CERN’s ECN3 hall, which exploits 400~GeV protons from the SPS. NA62 searches for di-lepton and hadronic final states have recently placed leading constraints~\cite{NA62:2023qyn,NA62:2023nhs,NA62:2025yzs}. The analysis of the full NA62 dataset in beam-dump mode is expected to be background-free. NA62 projections for the recorded $\sim 10^{18}$ protons on target (PoT) have been made publicly available~\cite{Jerhot:2936260,deBlas:2025gyz}.

The NA62 program is primarily focused on kaon physics and is currently expected to complete data-taking with the start of CERN’s LS3. For activities after LS3 at ECN3, two proposals were put forward: a mixed program of kaon and beam-dump physics (HIKE~\cite{HIKE:2023ext} with SHADOWS~\cite{Alviggi:2878470}), and an experiment at a dedicated beam-dump facility (SHiP~\cite{SHiP2023} at the BDF~\cite{Ahdida:2867743}).

With the approval of the beam-dump facility (BDF) and the SHiP experiment in March 2024, the ECN3 experimental hall is expected to be dedicated to the search for FIPs for the foreseeable future.

The originally proposed SHiP detector offers clear, leading sensitivity to FIP scenarios~\cite{Ahdida:2867743}, particularly for heavy neutral leptons (HNLs).
However, this detector configuration will not be realised immediately after LS3.
A down-scoped version of SHiP, using a helium balloon as the decay volume, is currently under consideration~\cite{Albanese:2948477}. 
By contrast, the fully operational NA62 detector is scheduled to begin dismantling in 2026.

In this document, we study different hypothetical detector configurations obtained by reconfiguring the currently operating NA62 detectors for the new BDF, and compare their physics reach to that of the originally proposed SHiP configuration for ECN3.
For each configuration, we uniformly assume $8 \times 10^{19}$ PoT delivered to the BDF,
which corresponds to about two years of data taking at the BDF, or roughly 80 times the statistics already collected by NA62.

For this study, it is instrumental to use a public tool in which all \textit{BC}s are implemented under unified theoretical assumptions~\cite{Dobrich:2024ajq}. One such tool is \textsc{Alpinist}~\cite{Schubert:2024hpm,Jerhot:2022chi}, which allows different geometries to be readily compared. The full software is publicly available at~\cite{jan_jerhot_2025_15526998}.

Using \textsc{Alpinist}, we explore the benefits and shortcomings of the considered detector configurations, showing that the production kinematics in different new-physics models imply that some \textit{BC}s are probed more efficiently than others.

Given the worldwide competition in the search for FIPs in the coming years~\cite{deBlas:2025gyz,Alimena:2025kjv}, we find that the parameter reach of these hypothetical scenarios is rather promising, under the well-motivated assumption that an essentially background-free search remains possible.

This paper is structured as follows. In Section~\ref{sec:BDF}, we review the nominal beam-line and dump configurations foreseen for ECN3. In Section~\ref{sec:detector_options}, we review the existing NA62 detectors located in ECN3 at the time of writing, as well as the originally proposed configuration for the SHiP experiment. We then describe possible hypothetical detector configurations with increasing levels of ``complexity", related to the degree of rearrangement required in the experimental hall. We assess the background limitations for these configurations in Section~\ref{sec:BKG}. We compare the physics reach of these configurations to that of the originally proposed SHiP detector layout in Section~\ref{sec:reach}. We conclude in Section~\ref{sec:conclusions}.

\section{Beam and the BDF target \label{sec:BDF}}
Before studying the impact of different detector configurations on the physics reach in ECN3, we briefly review the nominal beam and target foreseen for ECN3 after LS3.

A 400\,GeV proton beam will be slowly extracted from the SPS with a 1.2\,s flat-top spill and an intensity of $4 \times 10^{13}$~PoT per spill,\footnote{For comparison, NA62 in beam-dump mode received about $5 \times 10^{12}$~PoT per spill with 4.8\,s flat-top spills.} corresponding to an integrated intensity of $4 \times 10^{19}$~PoT per year delivered to ECN3, while maintaining an integrated intensity of $6 \times 10^{18}$~PoT per year to the other experimental areas, comparable to the pre-LS3 demand~\cite{Ahdida:2867743}. Assuming BDF commissioning by the end of 2031~\cite{Albanese:2948477}, an integrated intensity of $8 \times 10^{19}$~PoT could be delivered to ECN3 during the two years currently scheduled before LS4.

We assume the BDF design and geometry to be the same as described in \cite{Ahdida:2867743}:
\begin{itemize}
    \item The shielding of the target complex and the separation from ECN3 by a ventilation wall will allow access to ECN3 when the beam is stopped.
    \item The centre of the target is located in TCC8, 22\,m from the entrance of ECN3. The target is offset vertically by 50\,cm and the beam axis has a 11\,mrad horizontal angle with respect to the current T10 target used by NA62~\cite{Albanese:2948477}. With the target placed 133\,m downstream of T10, the RP issues which had to be addressed for HIKE~\cite{HIKE:2023ext} are assumed not to be a limitation in the following.
    \item The BDF target is composed of water-cooled blocks of titanium-zirconium-doped molybdenum alloy, cladded by a tantalum-alloy, in the core of the proton shower, followed by blocks of tantalum-cladded pure tungsten, together corresponding to 12 nuclear interaction lengths. Design improvements are being studied~\cite{SHiP:2025ows,Albanese:2948477}. Here, for simplicity, we assume molybdenum as the sole target material.
    \item The target complex is followed by a hadron absorber and a shield, sweeping the muons out of the acceptance of the detector located in ECN3.
    \item The intermediate detector designs proposed in the following section assume no background limitation for the FIP searches, or similar backgrounds, which are expected to be present with the SHiP detector design. More details on this point are discussed in Section~\ref{sec:BKG}.
\end{itemize}

\section{BDF detector options}
\label{sec:detector_options}

The NA62 experiment, currently installed and operational in ECN3, has released results from several searches for FIPs in the past few years~\cite{NA62:2023qyn,NA62:2023nhs,NA62:2025yzs}.
The aim of the following study is to investigate the physics potential of different hypothetical detector configurations obtained by reconfiguring the NA62 detectors for the new BDF, in order to maximise the physics output in ECN3 immediately after LS3. A brief description of the relevant NA62 detectors can be found in appendix ~\ref{app:na62}.

The BDF design discussed in the previous section is considered to be fixed, as well as the location of the proposed Scattering and Neutrino Detector (SND) just upstream of the decay volume, effectively requiring about 33\,m of space between the BDF target and the upstream edge of the decay volume. Therefore, in all that follows, the decay volume is assumed to be located 33.5\,m downstream of the centre of the BDF target. The current NA62 subdetectors upstream of the decay volume have to be removed, except for the veto hodoscope located at the upstream edge of the present NA62 decay volume, which can be moved downstream into ECN3. Possible modifications to the existing detectors are discussed in section~\ref{sec:modifications}.

Eight detector geometry configurations described in sections \ref{ssec:geo_0}, \ref{sec:geo_rearrange}, \ref{sec:geo_full} and appendices \ref{app:geo_a} and  \ref{app:geo_b} are considered and simulated with a toy MC as discussed in the next section. All configurations are assumed to be on-axis with respect to the incoming proton beam.
This can be achieved by a 12\,mrad kick of the incoming proton beam upstream of the target. Such an angular deviation effectively does not change the impact point and the shape of the shower in the 25\,cm wide and 144.5\,cm long target. 
For the minimal configuration, discussed further in section~\ref{ssec:geo_0}, we also consider the off-axis case, which corresponds to the nominal BDF conditions, with no dedicated beam re-steering. An overview of the main variables affecting the signal acceptance in the three configurations used in section \ref{sec:reach} is listed in table \ref{tab:geo_overview_part}. The full list can be found in table \ref{tab:geo_overview}.

\begin{table}[h!]
    \centering
    \begin{tabular}{l|c|c|c}
    \hline
      scenario & $l_\text{decay volume}\,[\text{m}]$ & $z_\text{ECal}\,[\text{m}]$ & $\Omega_\text{ECal}\,[\text{msr}]$ \\ \hline\hline
      \textbf{BDF 0} & 16.7 & 108.5 & 0.4 \\
      \textbf{BDF 3a} & 45 & 108.5 & 0.4 \\
      \textbf{BDF 4} & 50 & 93.6 & 2.74 \\
      \hline
    \end{tabular}
\caption{Overview of the main variables: the decay volume length, the distance between the target and the ECal and the solid angle covered by the ECal, which determine the detector acceptances for the configurations displayed in the physics reach in section \ref{sec:reach}.
    \label{tab:geo_overview_part}}
\end{table}

\subsection{General toy MC and signal-selection conditions}
\label{ssec:selection}
The FIP production and decay are simulated in the \textsc{Alpinist} framework~\cite{Jerhot:2022chi}. For the dark scalar production modes, 20M events are simulated for the $e^+e^-$ decay modes. For any other FIPs and decay modes, 1M events per mass bin are simulated for $e^+e^-(\nu)$ and $\mu^+\mu^-(\nu)$ final states and 500k events per mass bin for $2\gamma$, $2\pi$, $3\pi$, $4\pi$, $2\pi\gamma$, $2\pi\eta$, $2K$, $2K\pi$, $2K2\pi$, $\pi\ell$, $e\mu\nu$ final states.

For each considered final state, a specific signal-selection algorithm common to the detector configurations \textbf{BDF(0-3)} is applied and described below. Differences for the \textbf{BDF4} configuration are detailed in Section~\ref{sec:geo_full}.

\begin{itemize}
    \item[Photons:] All photons must be in the acceptance of an electromagnetic calorimeter (ECal) of octagonal shape, excluding a central ``hole'' with 8\,cm radius and must deposit at least 3\,GeV in total.

    For the case of a di-photon final state, the individual photon energy clusters in the ECal must be separated by at least 10\,cm, they must be at least 15\,cm away from the centre of the ECal and have an energy of at least 1\,GeV~ \cite{Dobrich:2019dxc}. The minimum energy condition is a limitation at the trigger level in order to avoid triggering by individual MIPs~\cite{NA62:2023qyn}, while the geometry conditions allow a good reconstruction of the individual electromagnetic clusters.
    
    \item[$e^+e^-(\nu)$:] The $e^\pm$ tracks must be in the acceptance of the first and last STRAW spectrometer stations. The STRAW stations are approximated as rotationally symmetric around the beam axis, with a radius of 1\,m and a central hole with a radius of 10\,cm. The same ECal selection criteria as for the two-$\gamma$ final states are applied.

    \item[$\mu^+\mu^-(\nu)$:] The $\mu^\pm$  tracks must be in the acceptance of the first and last STRAW stations and have momentum of at least 5\,GeV in order not to be kicked out of the acceptance of downstream detectors by the spectrometer magnet. Both tracks must also be in the acceptance of the muon veto detector (MUV3), excluding a central hole of 11\,cm radius.

    \item[$h^+h^-(\gamma)$:] Both hadronic tracks must be in the acceptance of the first and last STRAW stations and of the ECal and have momentum of at least 5\,GeV. Additional photons must be in the ECal acceptance and deposit at least 3\,GeV in total.

    \item[$e (h) \mu (\nu)$:] Both tracks must be in the acceptance of the first and last STRAW stations and of the ECal and must have energy of at least 5\,GeV. The electron/positron cluster must be at least 15\,cm from the ECal centre for a proper cluster reconstruction. The muon track must be in the acceptance of the muon veto.
\end{itemize}

\subsection{Minimal configuration (BDF 0)}
\label{ssec:geo_0}
In the minimal configuration, the current location of NA62 subdetectors is kept with no modification to the layout downstream of 33.5\,m from the centre of the BDF target, corresponding approximately to the location of the large-angle veto station 6. All detectors upstream of this point have to be removed from ECN3 and TCC8 to make space for the BDF target complex.

In this configuration, the vacuum tube has to be ended at 33.5\,m, which requires a new end cap with a larger diameter than the present one. Currently, section 15 of the vacuum tube after the large-angle veto station 6 starts at 33.155\,m from the centre of the BDF target, in principle (according to \cite{SHiP:2025ows} page 8) still allowing the space for placing of the SND detector and of the veto hodoscope upstream of the decay volume. The simulation of this configuration still assumes the decay volume to start at 33.5\,m. For final states including charged FIP daughter particles, the end of the decay volume corresponds to the first spectrometer station. For final states with only neutral daughter particles, the decay volume's endpoint is located at the ECal. The requirement of having at least a 10\,cm separation between the neutral particle positions at the ECal plane induces an effective reduction of the decay volume for neutral final states.

This configuration is also simulated with the option of the detector being off-axis.
The impact on the signal acceptance is discussed in section~\ref{sec:reach}.

\subsection{Configurations with detector rearrangements  (BDF 1 $\rightarrow$ 3)}
\label{sec:geo_rearrange}
From a geometrical perspective, the signal acceptance can be increased in two ways: by extending the decay volume and by increasing the angular acceptance. The impact of different choices is detailed in the appendices, here we describe the most sensitive configuration, denoted as \textbf{BDF 3a}.

The 17.69\,m long NA62 RICH detector can be removed at the cost of losing the $\pi$, $K$, and $p$ separation, which is currently possible in certain momentum ranges at NA62. The  NA62 STRAW spectrometer can then be accordingly placed closer to the calorimeter system, with the last STRAW station at approximately the location of the RICH mirror. Furthermore, we assume a possible 30\% reduction in the distances between the 4 STRAW stations, increasing the length of the decay volume by an additional 10.61\,m compared to the \textbf{BDF 0} configuration.

\begin{figure}[h!]
    \centering
\includegraphics[width=0.49\textwidth]{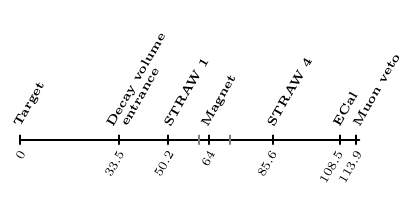}
\includegraphics[width=0.49\textwidth]{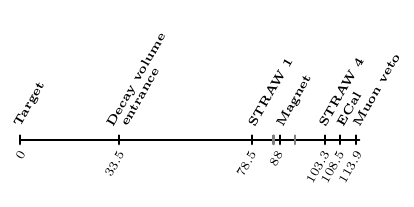}
    \caption{Location of the detectors and the decay volume along the $z$-axis with respect to the centre of the target in meters for \textbf{BDF 0} (left) and \textbf{BDF 3a} (right) configurations. The locations of the STRAW stations 2 and 3 are indicated in grey.}
    \label{fig:BDF03a}
\end{figure}

\begin{figure}[h!]
    \centering
\includegraphics[width=0.49\textwidth]{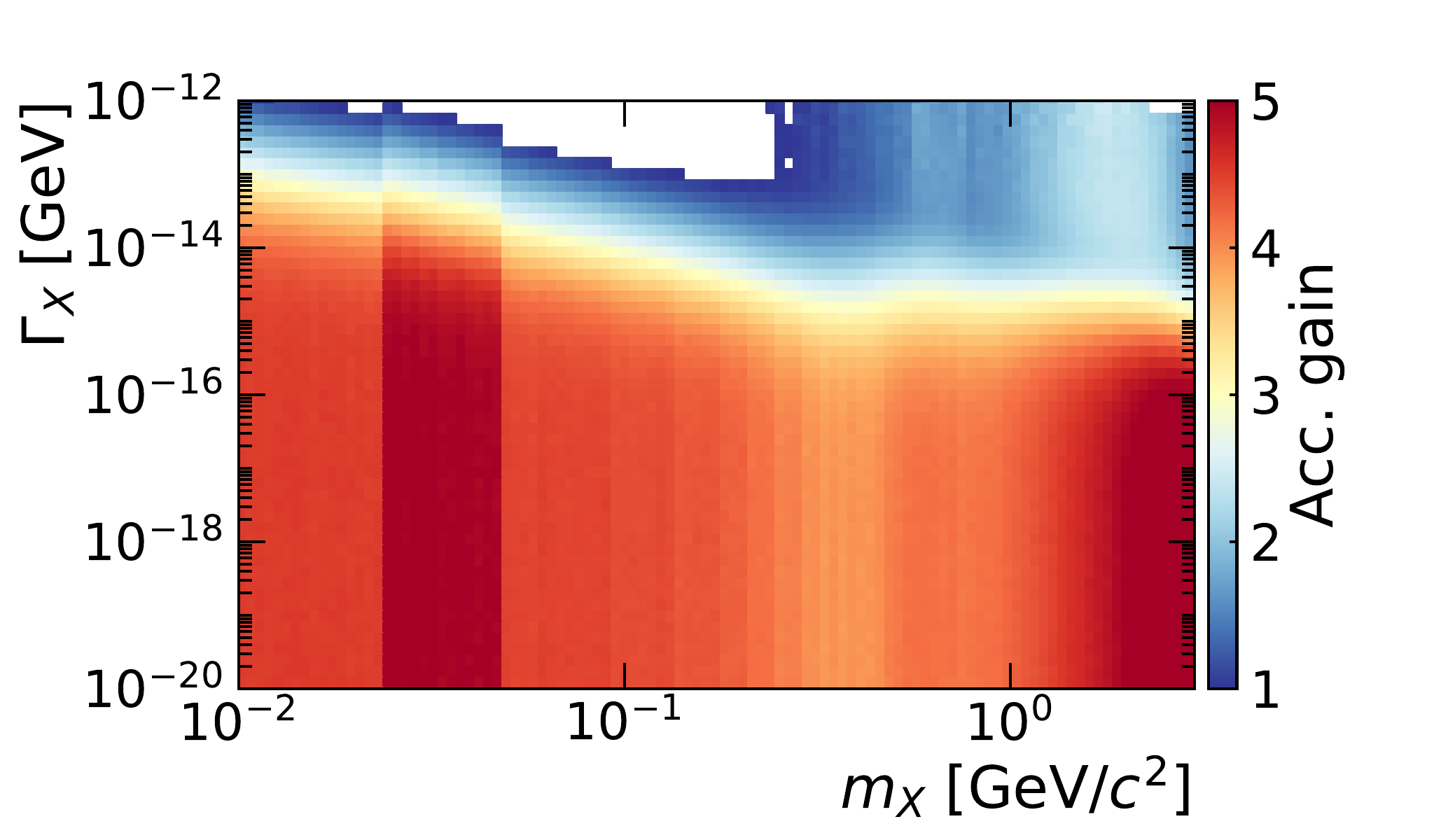}
\includegraphics[width=0.49\textwidth]{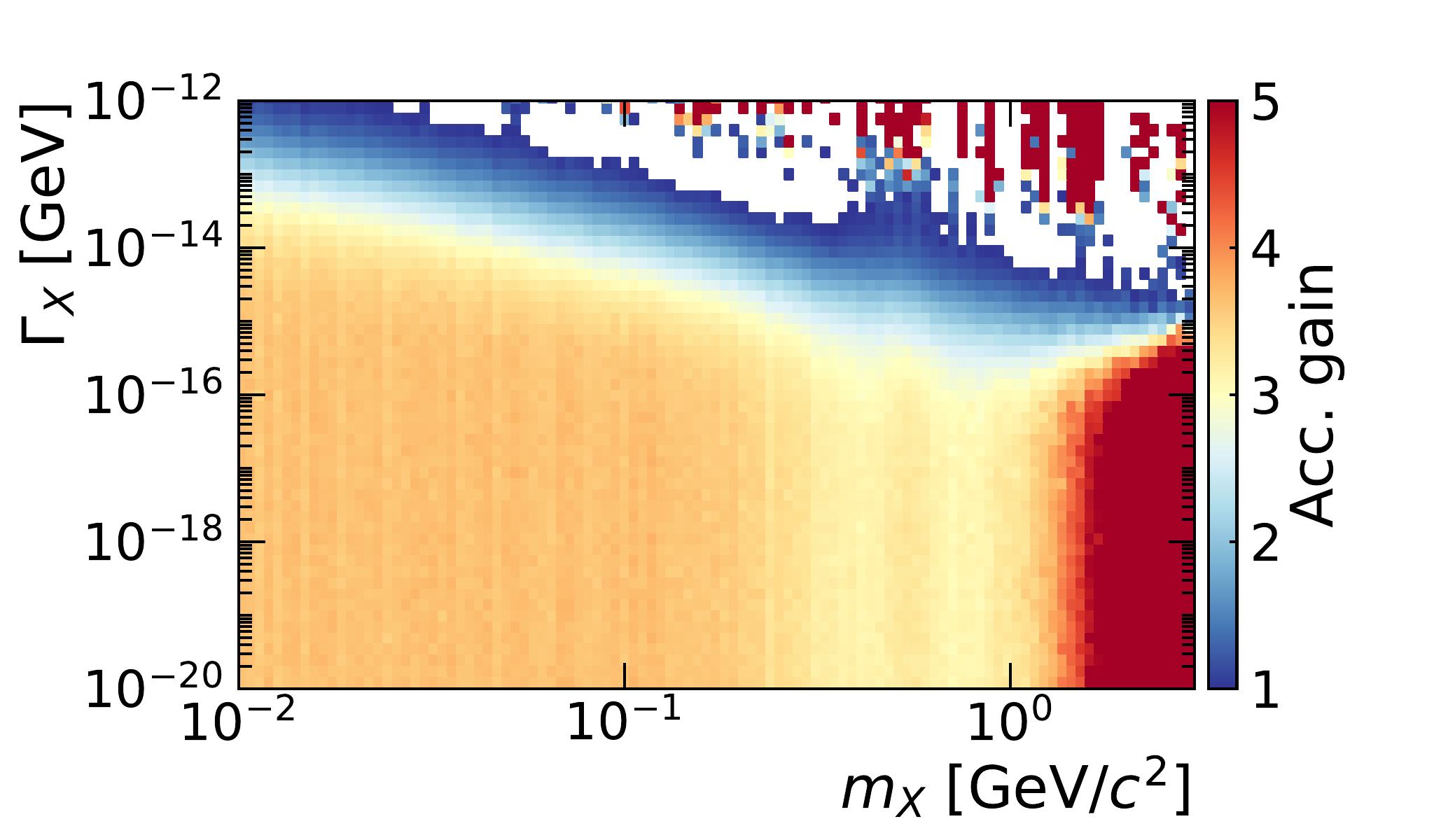}
    \caption{Ratio of the acceptance for setup \textbf{BDF 3a} to that for setup \textbf{BDF 0}, for a dark scalar decaying into $e^+ e^-$ produced by proton bremsstrahlung (left) and by \textit{B} meson decays (right).}
    \label{fig:bd3}
\end{figure}

The detector locations can be compared to the \textbf{BDF 0} layout in Fig.~\ref{fig:BDF03a}. The impact of the geometry adjustment on the signal yield for direct and meson-mediated production processes can be seen in Fig.~\ref{fig:bd3}.

\subsection{Modifying existing sub-detectors}
\label{sec:modifications}

A number of low-budget modifications to the existing NA62 detectors can be considered to further increase the signal and reduce the background yield. These modifications are not implemented in the toy MC, but here we outline several of them to indicate the direction in which the setup can be improved.

\begin{itemize}
    \item[\textbf{Upstream veto:}] In the current NA62 configuration, the upstream veto hodoscope is missing four central tiles for connecting the 75\,GeV beam-line with the decay volume. Closing this gap should be considered for the BDF operation to prevent halo particles from entering the decay volume undetected.

    \item[\textbf{Large-angle veto:}] With the whole detector placed significantly closer to the target compared to the current NA62 operated in the beam-dump mode, two possible sources of background should be considered:  halo particles entering the decay volume off-axis and decays in the decay volume with some particles escaping detection, leading to incomplete event reconstruction. These could be mitigated by putting the large-angle veto stations closer together. When keeping all 12 large-angle veto stations, the distances between the stations could be reduced by half. Such a modification would require cutting the current vacuum tube sections or constructing new ones. The impact of the distance between the last large-angle veto station and the calorimeter on the signal yield should be investigated.

    \item[\textbf{Non-instrumented parts of downstream detectors:}] With no beam passing through the centre of the experiment, parts of the detectors are not being instrumented, effectively reducing the signal acceptance. Namely, the centres of the STRAW stations, hodoscopes, calorimeters and muon veto. The non-instrumented centre of the calorimeters is currently considered in the toy MC simulations, while the impact on the STRAW track reconstruction efficiency is not known. These effects need to be evaluated, and the possibility of instrumenting some parts of the subdetectors is worth further investigation. For example, the upgrade of the scintillator tile detectors using the existing technology seems viable.

    \item[\textbf{Pre-shower detector:}] It is worth investigating if a pre-shower can be added to the calorimeter to add directionality, to address fully neutral final states better. Some understanding could be based on the EM showers from photons starting in the RICH mirror or in the hodoscopes, observed in the current kaon data. One of the possibilities would be a MicroMegas detector and reusing some of the readout of the current upstream detectors. Studies done for the pre-shower detector for the KLEVER Project \cite{KLEVERProject:2019aks} could be reused for the possible detector design.

\end{itemize}

\subsection{Full configuration (BDF 4)}
\label{sec:geo_full}

\begin{figure}[h!]
    \centering
    \begin{minipage}{0.49\textwidth}
        \includegraphics[width=0.9\textwidth]{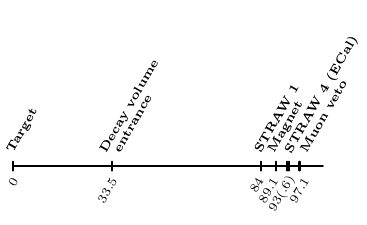}\\
    \end{minipage}
    \caption{Location of the detectors and the decay volume along the $z$-axis with respect to the centre of the target in meters for the \textbf{BDF 4} configuration.}
    \label{fig:BDF4}
\end{figure}

In the \textbf{BDF 4} configuration the geometric layout corresponds to the originally proposed SHiP design \cite{SHiP2023} with a 50\,m long decay volume followed by a spectrometer composed of four STRAW stations and ECal with transversal dimensions of $4 \times 6\,\text{m}^2$ and a magnet providing 0.65\,Tm vertical bending power. The detector layout is sketched in Fig.~\ref{fig:BDF4}.

The selection criteria applied are the same as for the other BDF configurations, except for the detector geometrical acceptances reflecting larger transversal dimensions without any holes in the detectors and the momenta of charged tracks, which are required to be $>1$\,GeV compared to the $>5$\,GeV cut applied above.

\section{Background assessment\label{sec:BKG}}
The background is assessed as follows: first, we extrapolate from the NA62 beam intensity to the BDF one, and we show that the background level is not inducing significant sensitivity limitations; then, we take into account the effect of the different dump-to-detector distances of the two layouts, and we argue that enough sweeping power can be installed to recover the NA62 conditions.

NA62 has demonstrated operation with negligible background. This holds (i) for final states with charged particles and closed kinematics, and (ii) for integrated proton rates well beyond $10^{18}$ PoT. The background has two components, which we will refer to as ``combinatorial'' and ``prompt'':
\begin{itemize}
    \item[\textbf{Combinatorial background:}] This component refers to random superposition of uncorrelated proton interaction products. It depends quadratically on the instantaneous intensity and is relevant only for final states with $\mu^+\mu^-$ pairs. The instantaneous beam intensity at the BDF is approximately a factor of 40 higher than that of the present NA62 beam-line. Extrapolating from the $10^{-2}$ background events expected in the NA62 $\mu^+\mu^-$ analysis at $1.4 \times 10^{17}$~PoT~\cite{NA62:2023qyn} to $8 \times 10^{19}$~PoT collected at the nominal BDF intensity, and applying the upstream veto---which has a measured single-particle inefficiency of 2\%~\cite{HIKE:2023ext} and was not used in ref.~\cite{NA62:2023qyn}---the expected number of combinatorial events for the NA62 geometry is $\mathcal{O}(10^{-1})$.

    \item[\textbf{Prompt background:}] This component refers to particles produced in secondary interactions of high-energy muons in the traversed material. It scales linearly with the integrated rate and is generally dominant for other final states than $\mu^+\mu^-$.  For the NA62 geometry, it was found that accounting for statistical scaling, this background is projected to be below 0.1 events for any final state with hadrons, and below 1 event for the $e^+e^-$ final state when restricting the search to invariant masses above 50~MeV (following the argument in ref.~\cite{HIKE:2023ext}).
\end{itemize}

We now consider the effect of the different geometrical setups when comparing NA62 and BDF. In the configurations shown in Fig.~\ref{fig:BDF03a}, the distance from the dump to the muon veto is $D_{\mu}^\mathrm{BDF0} = 114$~m, compared to $D_{\mu}^\mathrm{NA62} = 220$~m for NA62. Detailed simulations in ref.~\cite{Gatignon:2650989} of minimal modifications of the NA62 beam-line with $D_{\mu}^\mathrm{PBC} = 195$~m show only a modest increase of the muon rates with respect to the $D_{\mu}^\mathrm{NA62} = 220$~m setup. This is obtained with a sweeping system composed of three magnets of the MBPL (Magnet-Bending-from-PS Long) type (each with a field of 1.8~T and a length of 2~m)~\cite{Gatignon:711814}. Since in the \textbf{BDF~0} and \textbf{BDF~3a} configurations the distance is approximately two times smaller than that considered in ref.~\cite{Gatignon:2650989}, the corresponding single-muon rate is expected to increase by a similar factor. Therefore, to maintain the muon rates at the same level, an overall increase of the sweeping power by a factor of two or more is needed. 

There are good reasons to believe that a muon-sweeping system with sufficient rejection power to ensure a zero-background scenario in the considered setups\footnote{It must be emphasised that the sensitive detector area in these setups is roughly a factor of four smaller than that foreseen for SHiP.} can be achieved using conventional techniques. A complete background study and a full detector design would, of course, have to be carried out should the present document generate sufficient interest in the community. Here, however, we outline the main arguments supporting the assumption that a zero-background regime is achievable:

\begin{itemize}   
    \item Adding two MTR-type (Magnet-Target of Reduced gap) bending magnets (each with a field of 2~T and a length of 3.6~m)~\cite{Gatignon:711814} and three magnetised iron blocks (each with a field of 2~T and a length of 1.6~m) to the set of three MBPL magnets considered in ref.~\cite{Gatignon:2650989} would guarantee a total bending power increase by a factor of three, with a total nominal space of 18\,m.\footnote{This configuration is a minimal working example, only showing the approximate bending power achievable in a given space, rather than representing a concrete setup for the sweeping system.} 
    
    \item At the BDF, a conventional muon-sweeping system of this kind could be installed in the more than 20~m of space between the hadron absorber and the SND detector, designated for the muon shield~\cite{Albanese:2948477}, allowing further adjustments:
    indeed, muons of both charges must be swept, therefore the mutual distance and polarity of the various magnets must be carefully optimised; moreover, one must account for a minimum spacing between the magnets and for the effect of the return fields.

    \item The NA62 beam-dump configuration was only marginally optimised with respect to the muon rate at the muon veto, located 220~m downstream of the dump. For example, simulations indicate that NA62 could have achieved a $\times 10$ reduction in background above 15~GeV, at the cost of increasing the low-energy muon component, simply by retuning the currents in the two MTR-type bending magnets immediately downstream of the dump~\cite{Gatignon:2650989}.

\end{itemize}

\section{Physics reach \label{sec:reach}}

In the following, the sensitivity of the setup schemes described is estimated for the \textit{BC}s proposed in \cite{Beacham:2019nyx}. The sensitivity is expressed in terms of the expected 90\% confidence level exclusion bounds on the FIP model parameters (mass, coupling to the SM fields) in the absence of a new-physics signal. Three extreme cases are shown: \textbf{BDF 0}, the minimal-modification scheme in both on-axis (solid) and off-axis (dashed) configurations; \textbf{BDF 3a}, the scheme providing the largest sensitivity reach achievable with the NA62 detectors; \textbf{BDF 4}, the originally proposed SHiP configuration, as provided in ref.~\cite{SHiP2023}.

The same simulation with a toy MC is performed for all geometries for the same theory inputs and for the {\bf same PoT} to compare the performances on equal grounds. These are then compared with the expected NA62 reach simulated using the same toy MC and compared with the finalised analysis projections based on ref.~\cite{Jerhot:2936260}. For comparisons between \textsc{Alpinist} and sensitivities determined with \textsc{FairShip} see ref.~\cite{Schubert:2024hpm}.

\subsection{Dark photon}
A dark-photon field is considered, interacting with the SM through a kinetic-mixing coupling with the SM photon. The so-called \textit{BC1} scenario is characterised by the mixing parameter $\varepsilon$ and the dark photon mass $m_A^\prime$. The mixing parameter is introduced as $\mathcal{L} = - \varepsilon / ( 2 \cos \theta_{W})\, F^{\prime}_{\mu\nu}B^{\mu\nu}$, where $F^{\prime}_{\mu\nu}$ and $B_{\mu\nu}$ are the field strength tensors of the dark photon and the SM hypercharge gauge boson, respectively, and $\theta_{W}$ is the Weinberg angle. The calculation of the widths of individual dark photon decay channels is based on ref.~\cite{Ilten:2018crw}. 

\begin{figure}[t]
    \centering
\includegraphics[width=0.49\textwidth]{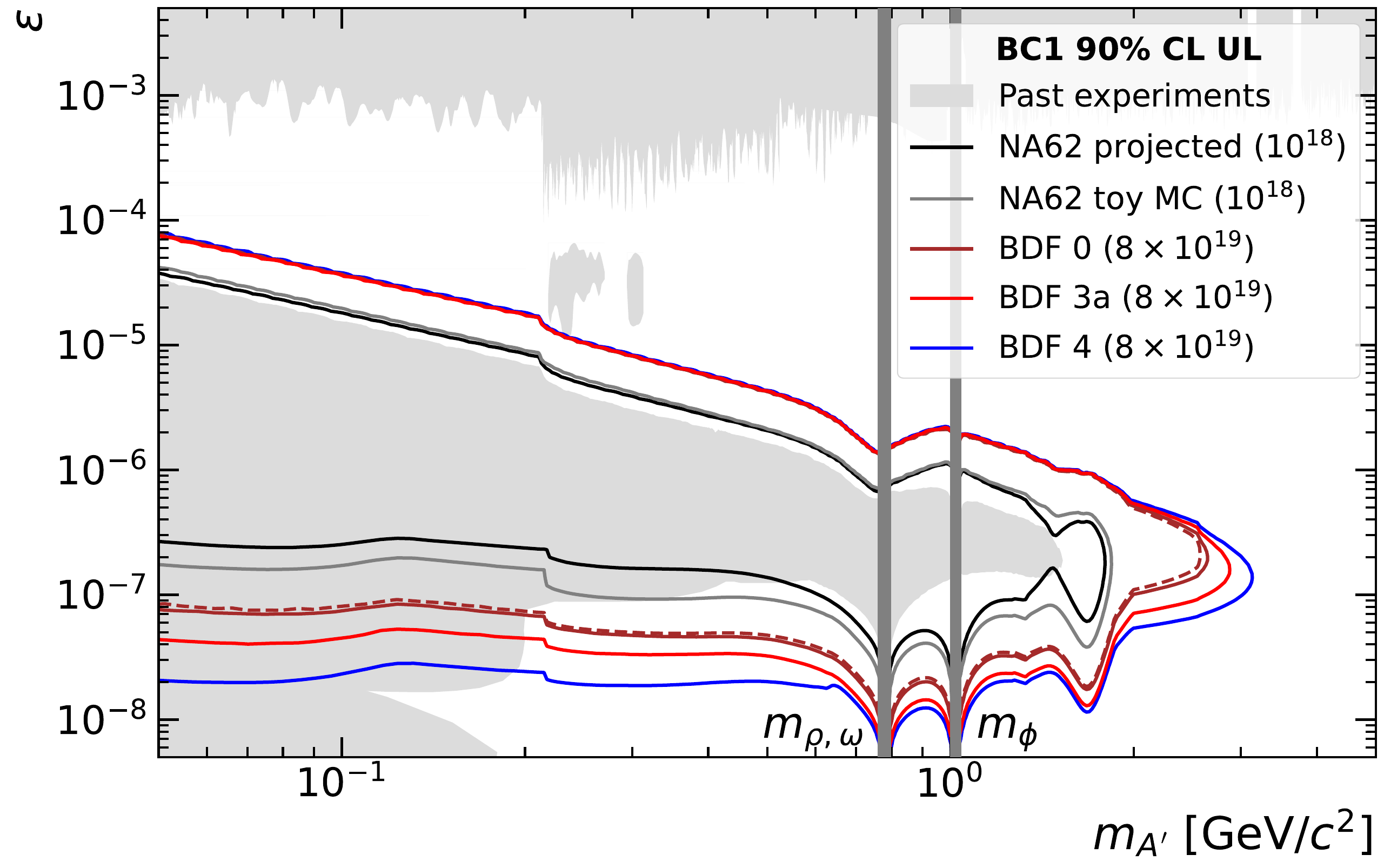}

    \caption{
     Exclusion regions at 90\% CL in the plane (mass, coupling) for the dark photon (\textit{BC1}) benchmark. In the absence of a FIP signal, the expected exclusion from NA62~\cite{NA62:2025yzs} at its final statistics~\cite{Jerhot:2936260} re-weighted to the updated Bremsstrahlung production is shown as a black line. For comparison, the grey line shows the expected sensitivity for the same statistics but estimated with the toy MC. The expected exclusions of the ``minimal post-LS3 scenario'' (\textbf{BDF 0}, dashed for off-axis, solid for on-axis) and the ``maximal post-LS3 scenario'' (\textbf{BDF 3a}) based on rearrangements of the present NA62 detectors can be compared to that of the originally proposed SHiP design (\textbf{BDF 4}) as brown, red and blue lines, respectively. Since the goal is to focus on the future sensitivity of experiments in ECN3, the sensitivity projections of DarkQuest~\cite{Apyan:2022tsd}, FASER~\cite{FASER:2022hcn}, LHCb~\cite{Gorkavenko:2023nbk} and other experiments operational in the relevant timescale are not displayed.}
    \label{fig:BC1}
\end{figure}

Compared to the projection from ref.~\cite{Jerhot:2936260}, which uses ref. \cite{Blumlein:2013cua}, here we use ref.~\cite{Foroughi-Abari:2024xlj} as a reference for the Bremsstrahlung production. For masses above 1.5\,GeV, also the production via Drell-Yan process is assumed, using the PDF set NNPDF 2.3 NLO ($\alpha_s(m_Z) = 0.118$)~\cite{Ball:2012cx}, with the events generated using MadGraph5\_aMC@NLO \cite{Alwall:2014hca} interfaced with \textsc{Pythia8.3}~\cite{Bierlich:2022pfr}. 

Figure \ref{fig:BC1} shows the resulting sensitivity of the various geometry setups. As in the high mass regime, the dark photon production is driven mostly by very forward processes, yet the difference between the on- and off-axis configurations is modest. The difference in the sensitivity between scenarios \textbf{BDF 0, 3a, 4} is relatively small and dominated by the length of the decay volume rather than the solid angle acceptance of the experiment.

\subsection{Higgs-like scalar}
The Higgs-like (dark) scalar interacts with the SM Higgs doublet as $\mathcal{L} = - (\mu_S S + \lambda_S S^2 ) \, H^{\dag} H$. Below the EW symmetry breaking scale, the dark scalar mixes with the SM Higgs boson $h$ with a mixing angle $\theta \simeq \mu_S v / (m_h^2 - m_S^2)$, where $v$ is the vacuum expectation value of the Higgs field. The quartic coupling $\lambda_S$ affects the pair production of the dark scalars. Two \textit{BC} scenarios are considered:
\begin{itemize}
    \item \textit{BC4}: $\lambda_S = 0$, therefore, there is no signal enhancement from the $B \to K SS$ decay.
    \item \textit{BC5}: $\lambda_S = 5 \times 10^{-4}$, corresponding to the limit on the branching ratio for the Higgs boson decay into a pair of dark scalars $\mathrm{BR}_{h \to SS} = 10^{-2}$ projected to be achievable at HL-LHC \cite{Dainese:2703572}; in this assumption, the rate for $B \to K SS$ decays is enhanced at very low values of $\theta$ dominating over the $S$ production from $B \to K S$ decays and proton bremsstrahlung.
\end{itemize}

\begin{figure}[t]
    \centering
\includegraphics[width=0.49\textwidth]{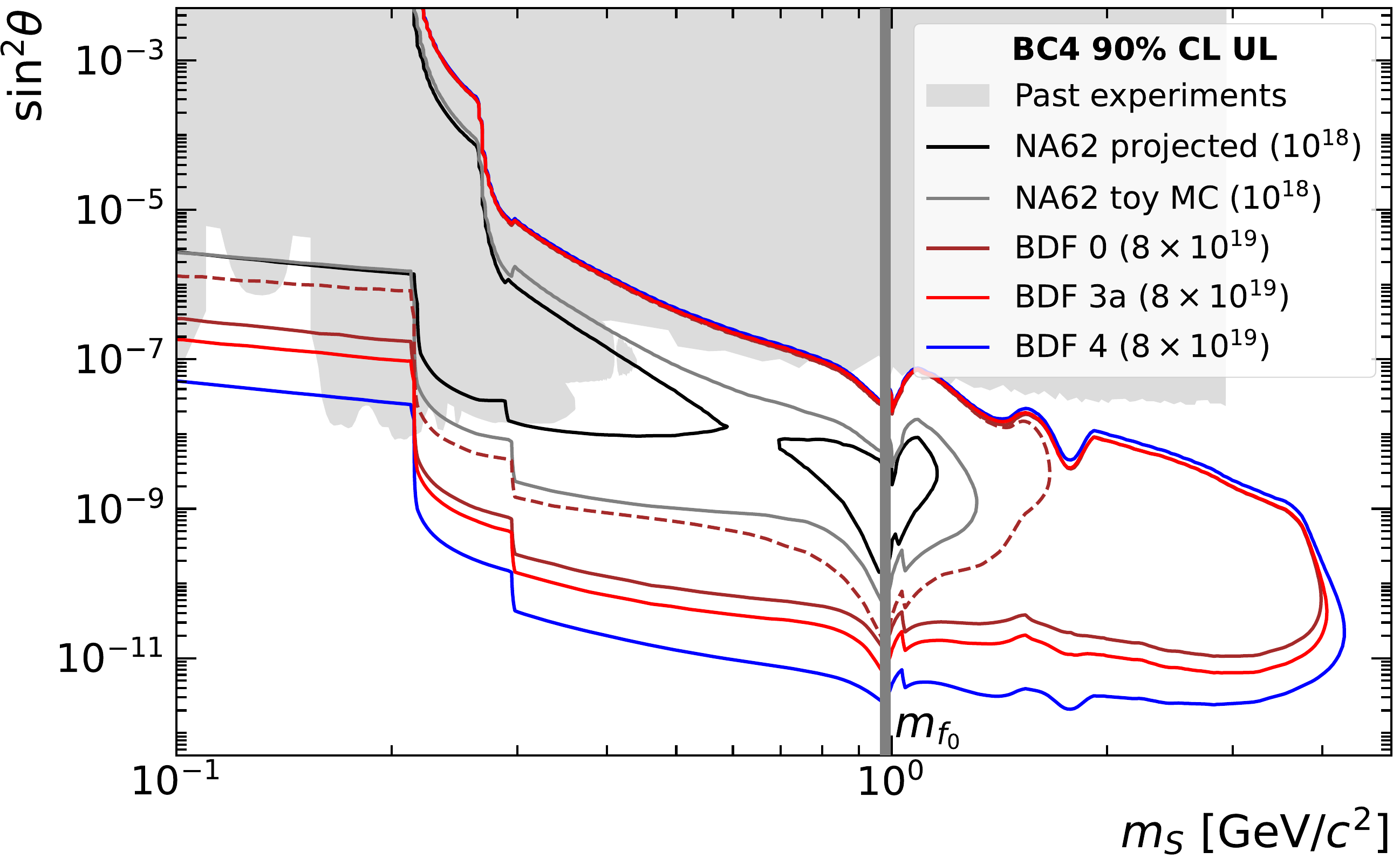}
\includegraphics[width=0.49\textwidth]{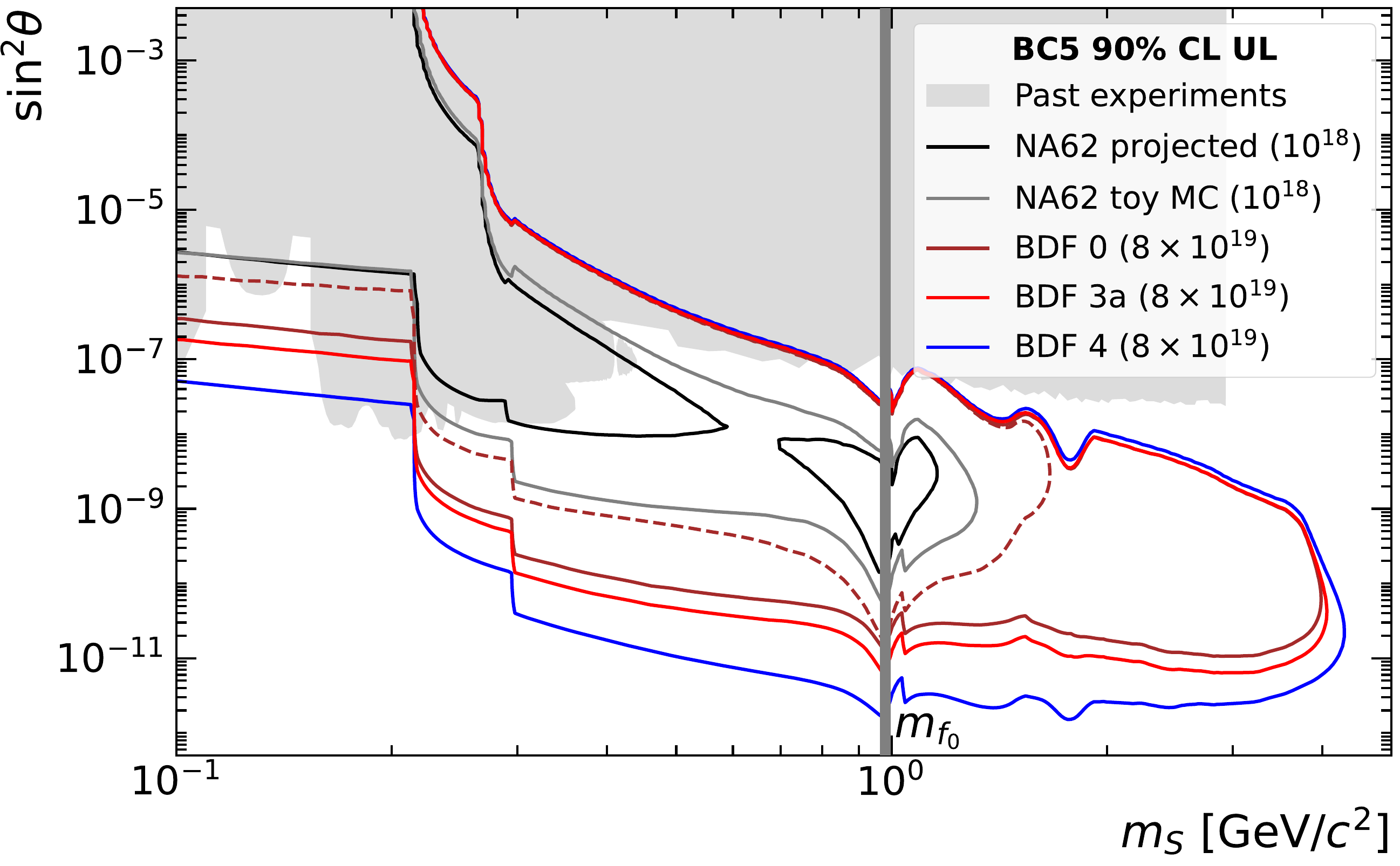}
    \caption{Exclusion regions at 90\% CL in the plane (mass, coupling) for the Higgs-like scalar (\textit{BC4} and \textit{BC5}) benchmarks. For details, see the caption of Fig.~\ref{fig:BC1}.}
    \label{fig:BC45}
\end{figure}

Figure \ref{fig:BC45} shows the sensitivity comparison for the two benchmarks. Similarly to the dark photon case, the difference between the sensitivity of the various detector setups for post-LS3 scenarios is marginal, with a noticeable difference for the off-axis configuration (see Appendix~\ref{app:brems} for more details). The production is indeed dominated by very forward processes such as Bremsstrahlung, where the derivation from ref.~\cite{Foroughi-Abari:2021zbm} is adopted in the simulation.

Compared to~\cite{Jerhot:2936260} the $2\pi$ and $2K$ decay widths below 2\,GeV mass are calculated using ref.~\cite{Blackstone:2024ouf}, including a recast of the projected sensitivity of NA62 from the published analysis. The hadronic widths above 2\,GeV mass, including the $4\pi$ and $2K2\pi$ channels, are evaluated using the approach described in ref.~\cite{Gieseke:2025gfq}. The total hadronic width and the $4\pi$ and $2K2\pi$ widths below 2\,GeV are calculated by applying the method of ref.~\cite{Gieseke:2025gfq} also below the 2\,GeV mass and forcing continuity with the result of ref.~\cite{Blackstone:2024ouf}. Compared to the 2-track analyses projections, the toy MC exclusion regions include 4-track final states as well as fully neutral ones such as $2\pi^0$ or $4\pi^0$.

\begin{figure}[t]
    \centering
\includegraphics[width=0.49\textwidth]{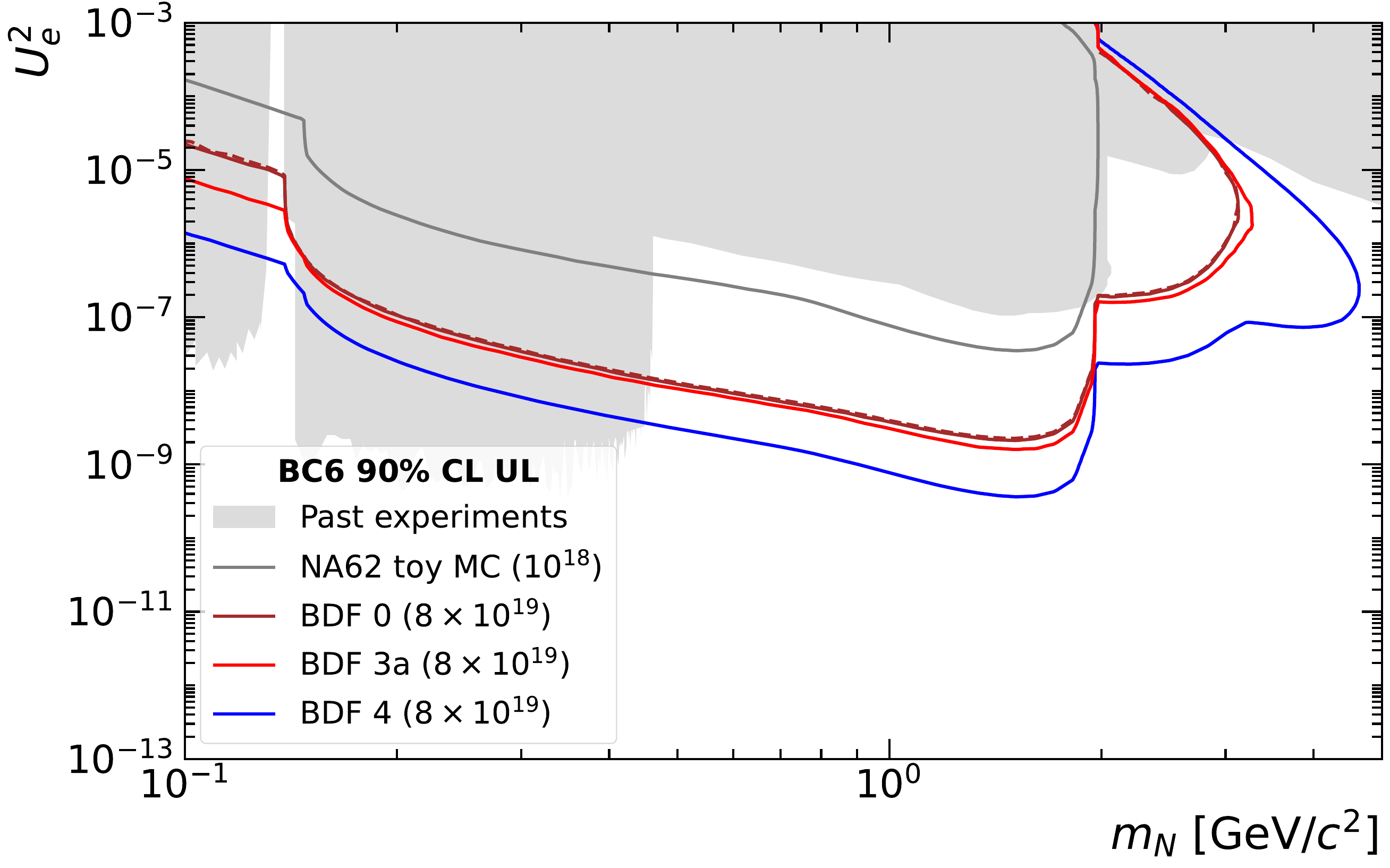}
\includegraphics[width=0.49\textwidth]{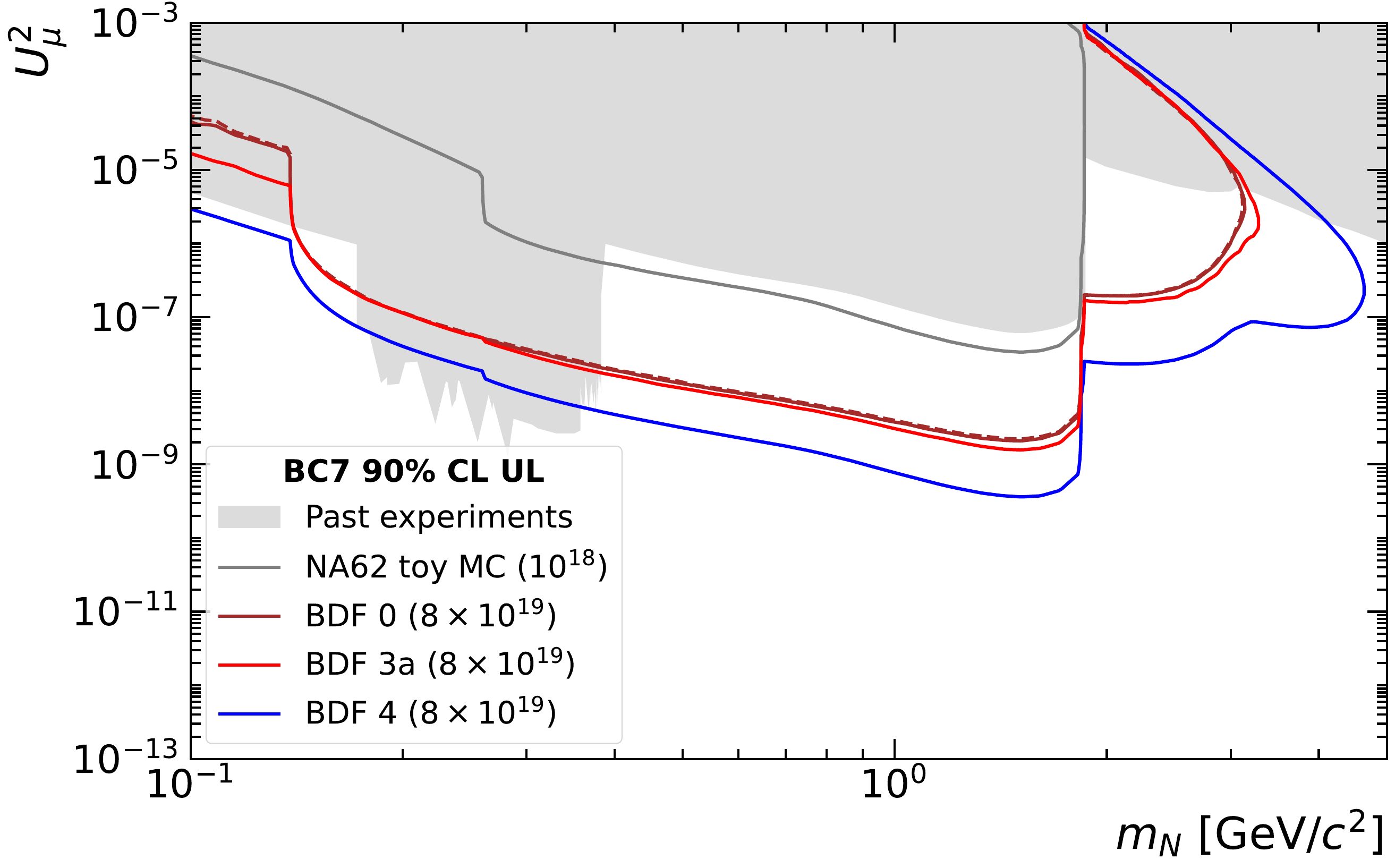}
\includegraphics[width=0.49\textwidth]{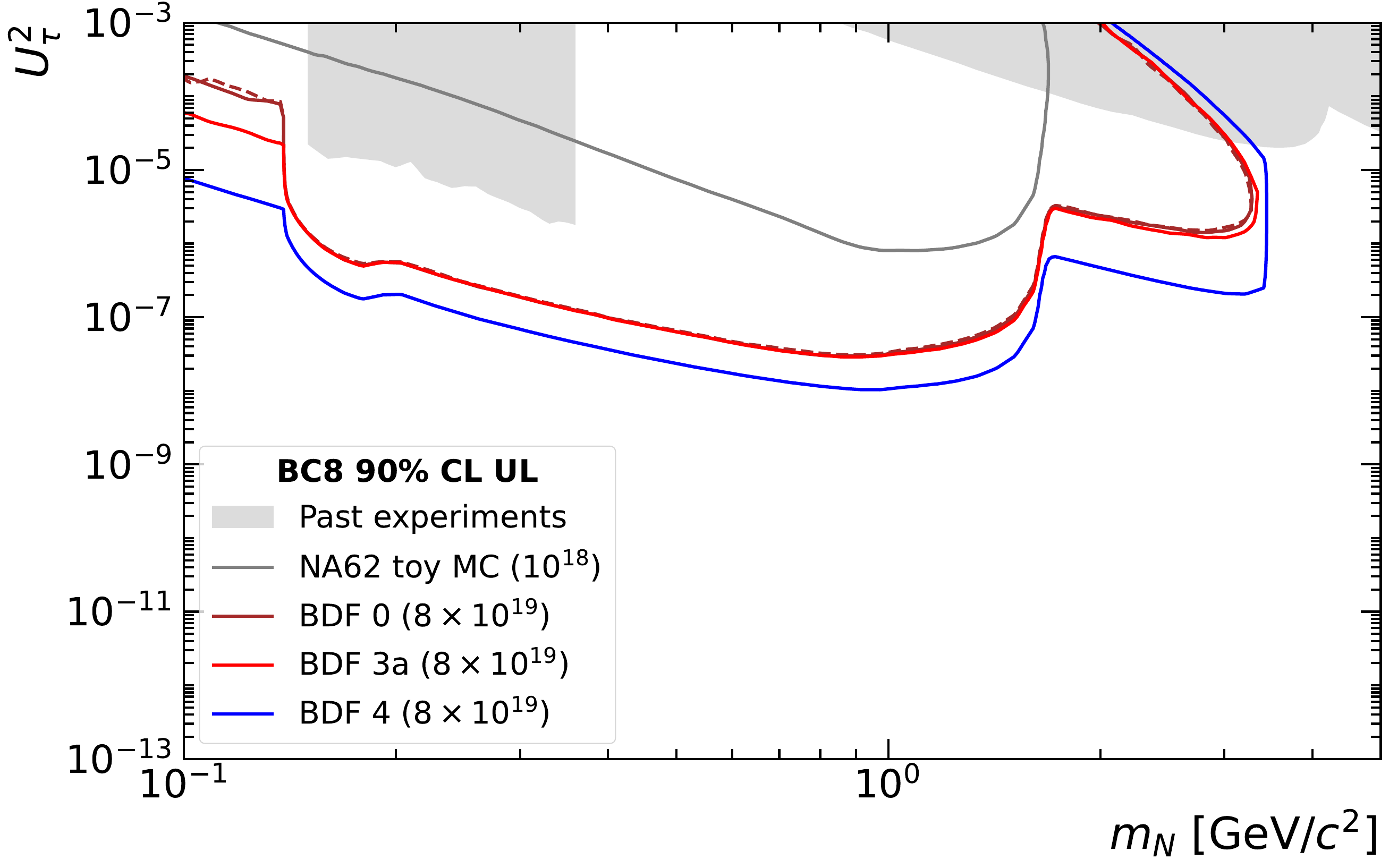}
    \caption{Exclusion regions at 90\% CL in the plane (mass, coupling) for the electron-coupled (\textit{BC6}, top left panel), muon-coupled (\textit{BC7}, top right panel) and tau-coupled  (\textit{BC8}, bottom panel) HNL benchmarks.  For details, see the caption of Fig.~\ref{fig:BC1}.}
    \label{fig:BC678}
    \vspace{4mm}
\end{figure}

\subsection{Heavy neutral leptons}
Heavy neutral leptons (HNLs) $N_I$ couple to the Higgs doublet and SM lepton doublet $L_\alpha$ as $\mathcal{L} \supset -i\sum_{\alpha, I} F_{\alpha I}\bar{L}_\alpha\sigma_2 H^* N_I$, where $F$ is the Yukawa coupling. The resulting mixing with the SM neutrinos is parametrised by mixing matrix elements $U_{\alpha I}\simeq |F_{\alpha I}| v / (\sqrt{2} m_{N I})$. Three \textit{BC} scenarios are considered, each one assuming a single HNL with coupling $U_\ell$ to a single lepton generation $\ell \in \left\lbrace e, \mu, \tau \right\rbrace$. \textit{BC6}: $U_e \neq 0$ and $U_\mu = U_\tau = 0$ and similarly for \textit{BC7} and \textit{BC8}.

The HNLs are produced in decays of \textit{D} and \textit{B} mesons. We use the results of ref.~\cite{Schubert:2024hpm} for the kinematic spectra of these mesons at small centre of mass energies. The expected sensitivities for the three benchmarks are displayed in Fig.~\ref{fig:BC678}. Here, the solid angle accepted for the signal is of relevance, significantly favouring \textbf{BDF 4} above the \textit{D} meson mass threshold with respect to the other ECN3 setup schemes considered. The off- versus on-axis configuration, as well as the length of the decay volume, which significantly varies between scenarios \textbf{BDF 0} and \textbf{BDF 3a}, have a negligible impact.

\begin{figure}[t]
    \centering
\includegraphics[width=0.48\textwidth]{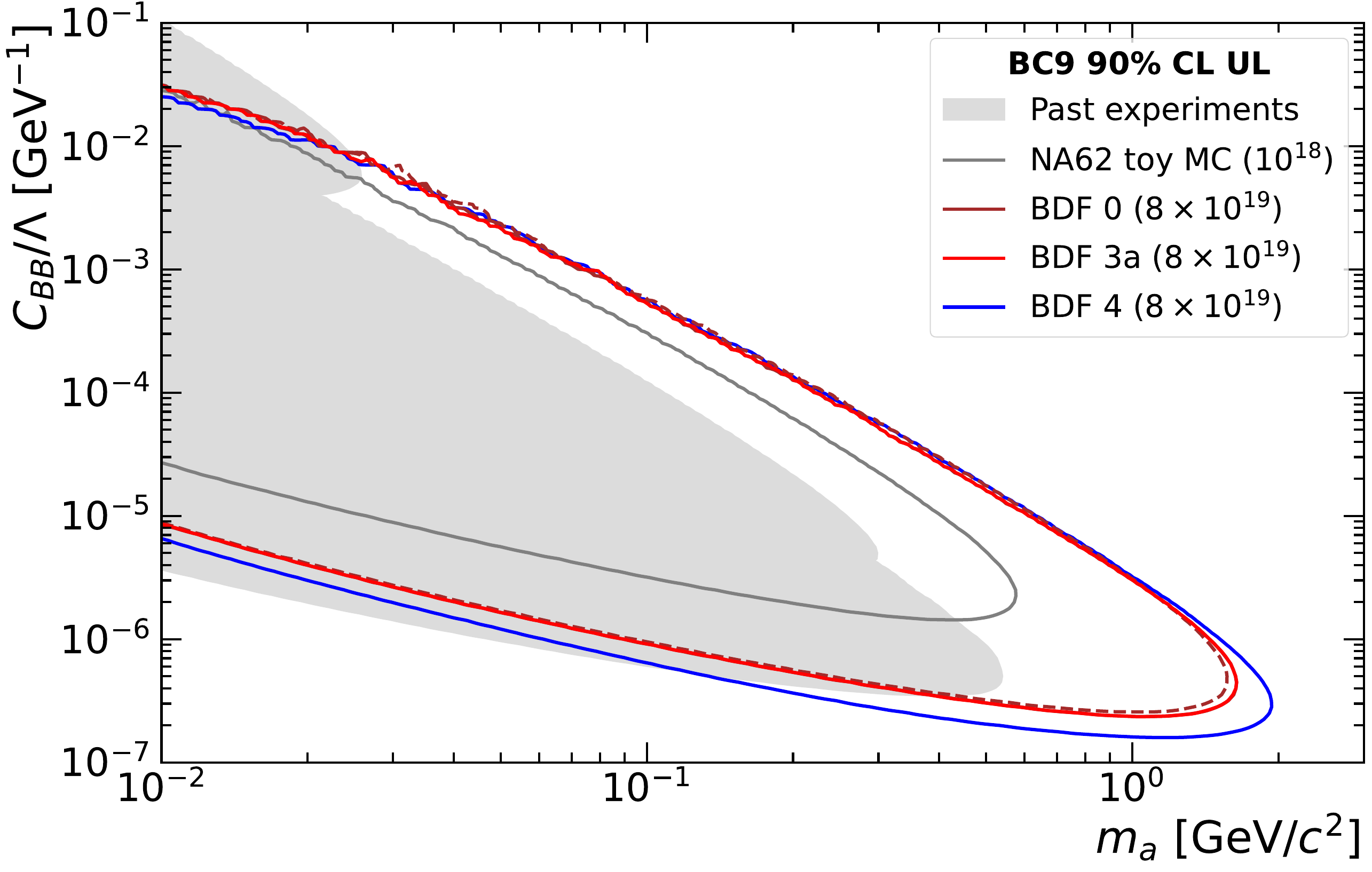}
\includegraphics[width=0.51\textwidth]{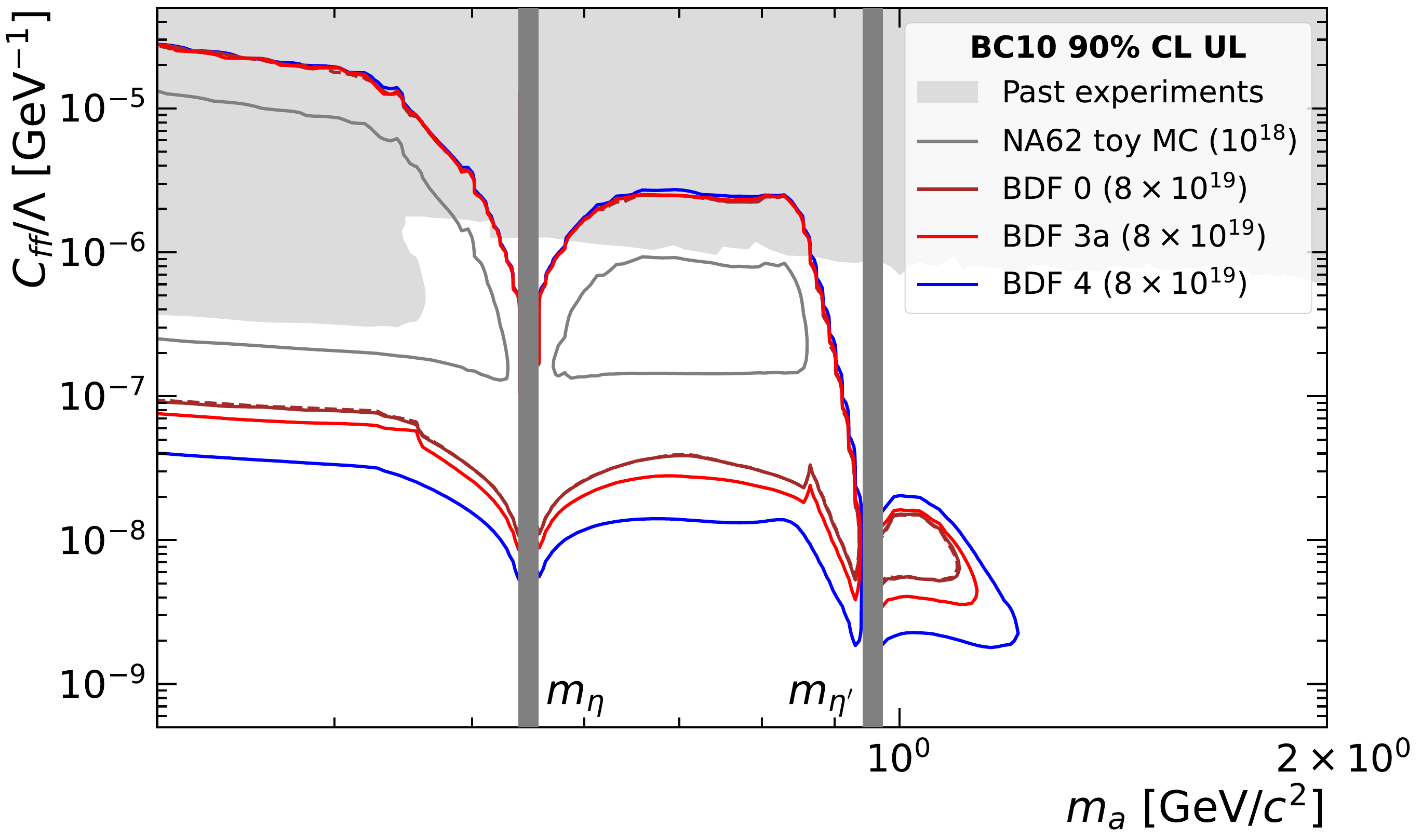}
\includegraphics[width=0.49\textwidth]{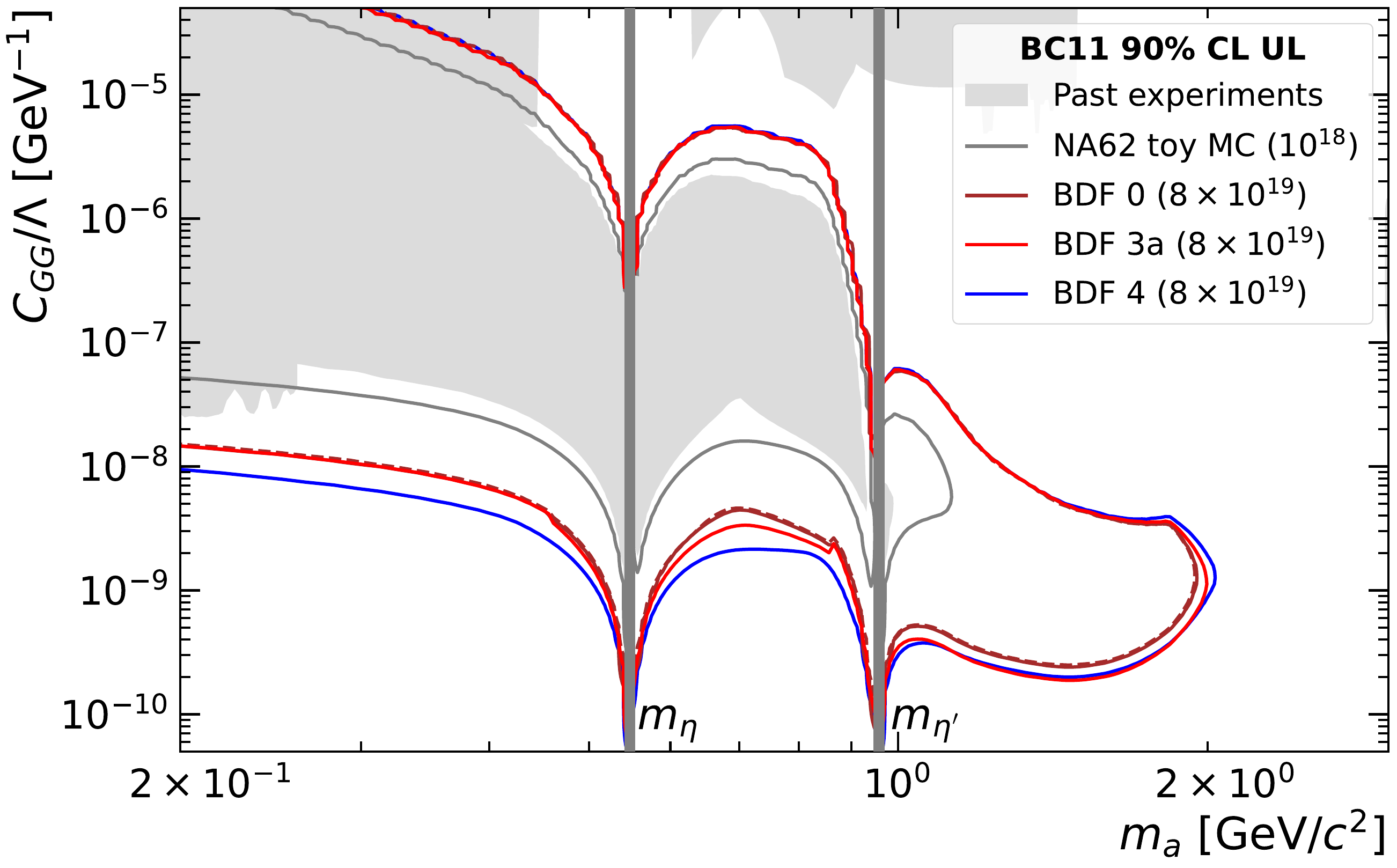}
    \caption{Exclusion regions at 90\% CL in the plane (mass, coupling) for the photon-coupled (\textit{BC9}, top left panel), fermion-coupled (\textit{BC10}, top right panel) and gluon-coupled  (\textit{BC11}, bottom panel) axion-like particle benchmarks. The NA62 sensitivity to the di-photon, di-lepton and hadronic final states with a full statistics of $10^{18}$ PoT in the \textit{BC9}--\textit{BC11} models is shown as the grey contour. The expected exclusion bounds in the absence of a FIP signal for the ECN3 post-LS3 setups \textbf{BDF 0, 3a, 4} are shown as the brown (dashed for off-axis, solid for on-axis), red, and blue contours, respectively. For all the evaluations, $\Lambda = 1\,\mathrm{TeV}$ is assumed.}
    \label{fig:BC91011}
\end{figure}

\subsection{Axion-like particles}
An axion-like particle $a$ can couple to the SM fermions $f$ and gauge bosons $V$, respectively: $\mathcal{L} = C_{ff}/(2\Lambda) \; \partial_{\mu}  a \bar{f}\gamma^{\mu}\gamma^5 f$ and $\mathcal{L} =  g^2(C_{VV}/\Lambda)\, a V_{\mu\nu}\tilde{V}^{\mu\nu}$, where $\Lambda$ is the UV scale. In the following, we assume $\Lambda = 1\,\mathrm{TeV}$. Three \textit{BC} scenarios are considered:
\begin{itemize}
    \item \textit{BC9}: ALP coupled to photons (assuming coupling to the SM $B$ boson at $\Lambda = 1\,\mathrm{TeV}$), produced via Primakoff process and decaying into two photons;
    \item \textit{BC10}: Fermion-coupled ALPs, decaying into two leptons and hadrons;
    \item \textit{BC11}: Gluon-coupled ALPs, decaying into two photons and hadrons.
\end{itemize}

Figure \ref{fig:BC91011} shows the sensitivity for the ALPs in the different benchmarks. ALPs produced in \textit{BC9} and \textit{BC11} tend to be forward emitted; therefore, the difference between the various ECN3 geometries is tiny, resembling the cases of \textit{BC1} and \textit{BC4}. In the \textit{BC10} model, the decays of \textit{B} mesons play a stronger role in ALP production, favouring a larger solid angle geometry.

\section{Conclusion \label{sec:conclusions}}

The search for FIPs has gained a lot of traction in the past decade: weakly coupled particles at the MeV-GeV scale are theoretically motivated and can be probed in medium-size experiments exploiting existing accelerator infrastructures. Experiments in this field can be realised on a time scale that makes them also an interesting ``bridge'' in a HEP context that is envisioning the next accelerator project.

Examples of present beam-dump experiments are the CERN experiments such as NA62 run in beam-dump mode, and NA64 \cite{NA64:2025ddk}, which have released a number of results in the past few years. Novel, dedicated experiments elsewhere are upcoming, and a number of collider-based experiments are expected to also search large FIP parameter regions, see \cite{Alimena:2025kjv} for a recent review.

One of the most powerful facilities for probing the parameter space of FIPs will be the BDF, located in the ECN3 hall at CERN’s Prevessin site. With $\sim 4 \times 10^{19}$ dumped protons per year at an energy of 400 GeV, BDF will open the possibility to probe extremely weak couplings. The SHiP experiment and its accompanying muon sweeping system are an ambitious project to fully exploit the opportunities provided by BDF. However, the originally proposed version of SHiP is expected to become operational later than the BDF~\cite{Albanese:2948477}.

In this letter, we have explored different scenarios for detector set-ups that could exploit the BDF's physics potential and be ready very soon after LS3: we have shown that even the most minimalistic modification of the existing NA62 experiment's detectors can, in many benchmark scenarios, already provide a very competitive sensitivity. We have based these estimates on the public tool \textsc{Alpinist}, which ensures equal assumptions for production of FIPs in terms of cross-section, kinematics and other theoretical inputs for all scenarios under scrutiny.

We caution that all estimates shown are assuming a zero-background scenario. The background has been proven to be completely negligible at NA62 for di-leptonic and a number of hadronic final states~\cite{NA62:2023qyn,NA62:2023nhs,NA62:2025yzs}.  Given the transversal size of the NA62 detector and results from studies in the context of the PBC, we have detailed why a zero-background scenario is achievable at the BDF as well.

To maximally exploit the protons deliverable to ECN3 after LS3, a detector ready immediately after LS3 or when commissioning the BDF would be the best-case scenario. With this study, we have shown that with the existing detectors of the NA62 experiment, even when left simply in place downstream of the BDF, a sizeable parameter space of FIPs could be probed.

\section*{Acknowledgements}
We thank F. Brizioli, N. Doble, L. Gatignon,  C. Lazzeroni, M. Raggi and G. Ruggiero for comments on the manuscript draft, and numerous NA62 colleagues for discussions.
BD and JJ acknowledge funding through the European Research Council under grant ERC-2018-StG-802836 (AxScale).

\bibliographystyle{JHEP_improved}
\bibliography{bibliography}

\newpage

\appendix

\section{The NA62 detectors}
\label{app:na62}
Here we provide a short description of the NA62 detectors mentioned in the main text. For a full description, see \cite{NA62:2017rwk} and references therein.
\begin{itemize}
    \item Upstream veto (ANTI0): The ANTI0 hodoscope is located immediately upstream of the vacuum tank. In order to cover the whole geometrical acceptance of the first section of the vacuum vessel it has an octagonal shape encompassing a circle with diameter of 2160\,mm. A central hole with dimensions $232\times232\,\mathrm{mm}^2$ is left not instrumented to allow passage for the beam pipe. It is composed of $124\times124\,\mathrm{mm}^2$ scintillating tiles arranged in a chessboard-format.
    \item Large angle veto (LAV): For vetoing photons emitted in the decay volume at angles between 8.5 and 50\,mrad with respect to the beam line, a veto system consisting of 12 ring-shaped stations is placed along the whole downstream region. Each station has four to five rows of lead glass blocks placed around the perimeter of the volume inside. Electromagnetic showers developing in the lead glass are detected via the Cherenkov light.
    \item STRAW: The STRAW spectrometer, occupying the last 35\,m of the NA62 vacuum tank (total of 117\,m), is composed of four chambers, two located before and two after a large aperture dipole magnet (MNP33). Each chamber has four, so-called `views' perpendicular to the beam axis under different angles with a 12\,cm gap for the beam left near the centre. Each view consists of 448 straw tubes made of a very thin ($36\,\mu\mathrm{m}$) PET foils. Each tube is 2160\,mm long and about 9.82\,mm in diameter. The single straw tubes are filled with a mixture of 30\% of CO2 and 70\% of argon at atmospheric pressure. 
    \item RICH: The Ring Imaging Cherenkov Counter (RICH) 
    is composed of a 17.5 m long vessel with a conical shape with a diameter of 3.2\,m at the downstream and 4\,m at the upstream end. 
    It is filled with gaseous neon at a pressure of 990\,mbar. 
    \item ECal (LKr): The Liquid Krypton calorimeter (LKr) is an electromagnetic calorimeter. It is quasi-homogeneous and placed inside a cryostat filled with $9\,\mathrm{m}^3$ of liquid krypton. The outer radius of the LKr is around 128\,cm. The inner 8\,cm are left for the beam pipe. The active area of the LKr extends for about 127\,cm in depth, corresponding to $27\,X_0$ and allowing a full containment of the electromagnetic showers.
    \item Muon veto (MUV): The muon veto (MUV) is a system of three independent detectors behind the LKr: two hadronic calorimeters called MUV1 and MUV2 and a hodoscope called MUV3. MUV1 and MUV2, covering a plane $260\times260\,\mathrm{cm}^2$, are iron-scintillator sandwich calorimeters of twelve vertical and twelve horizontal layers of iron, corresponding to about $\sim8 X_0$, which alternate with scintillator strips. An iron absorber of 80\,cm is placed between MUV2 and MUV3. The latter is composed of $22\times22\,\mathrm{cm}$ scintillating tiles. 
\end{itemize}

\section{Fixed angular acceptance configurations \textbf{BDF 1a, 2a, 3a}}
\label{app:geo_a}

With the ECal located as in configuration 0, the signal yield can be increased by extending the decay volume through a detector rearrangement.
Three configurations are considered:
\begin{itemize}
    \item 1a) The distance between the last two STRAW spectrometer stations is reduced by approximately 30\%. Consequently, the STRAW stations 1, 2 and 3, as well as the spectrometer magnet, are shifted downstream by approximately 3.9\,m so that the distances between the first two and last two STRAW spectrometer stations are equal to approximately 10.5\,m. The location of the other subdetectors remains unchanged.
    
\begin{figure}[h!]
    \centering
\includegraphics[width=0.49\textwidth]{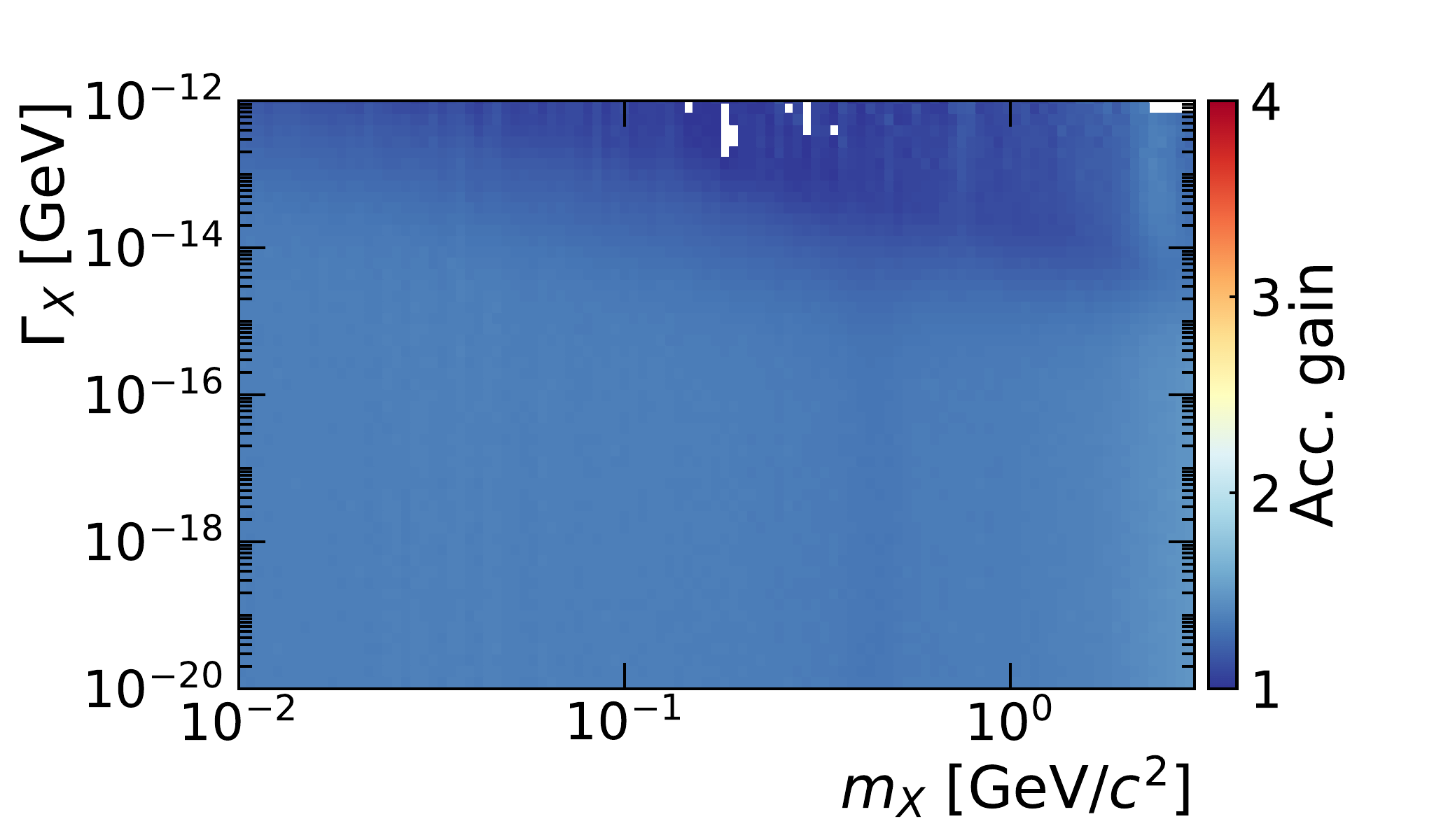}
\includegraphics[width=0.49\textwidth]{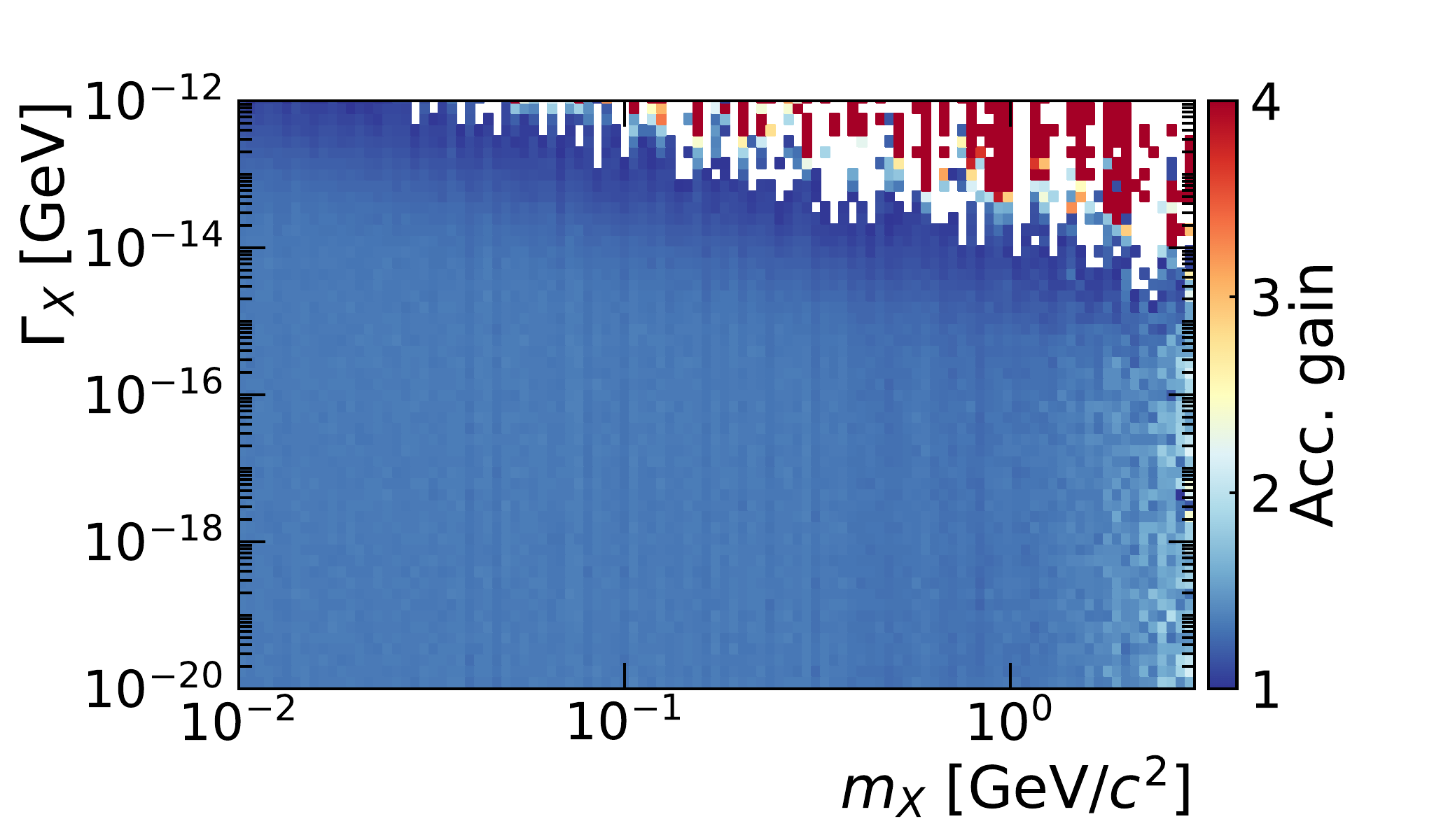}
    \caption{Acceptance increase in \textbf{BDF 1a} compared to \textbf{BDF 0} for dark scalar decaying into $e^+ e^-$ produced in proton bremsstrahlung (left) and in \textit{B} meson decays (right).}
    \label{fig:bd1a}
\end{figure}

    \item 2a) The distances between the 4 STRAW stations are reduced by 30\%, effectively increasing the size of the decay volume by 10.613\,m.  The detector locations can be compared to the \textbf{BDF 1a} layout in Fig.\ref{fig:BDF12a}.

\begin{figure}[h!]
    \centering
\includegraphics[width=0.49\textwidth]{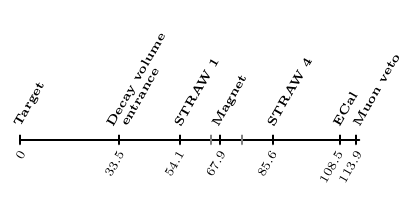}
\includegraphics[width=0.49\textwidth]{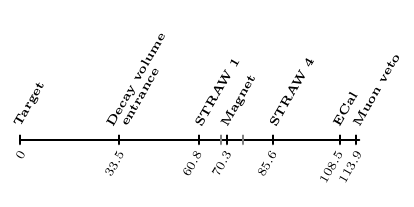}
    \caption{Location of the detectors and the decay volume along the $z$-axis with respect to the centre of the target in meters for \textbf{BDF 1a} (left) and \textbf{BDF 2a} (right) configurations. The locations of the STRAW stations 2 and 3 are indicated in grey.}
    \label{fig:BDF12a}
\end{figure}

\begin{figure}[h!]
    \centering
\includegraphics[width=0.49\textwidth]{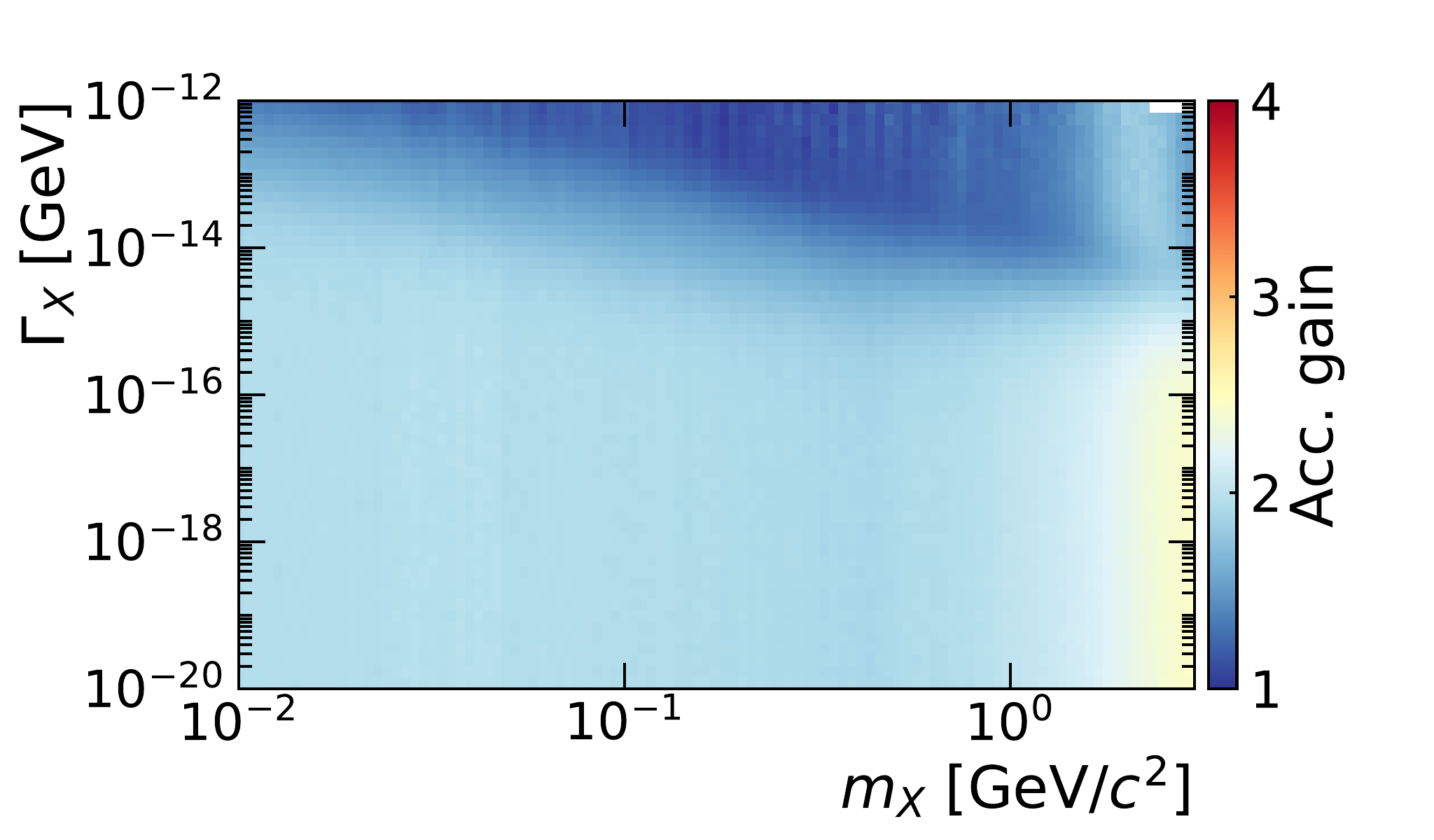}
\includegraphics[width=0.49\textwidth]{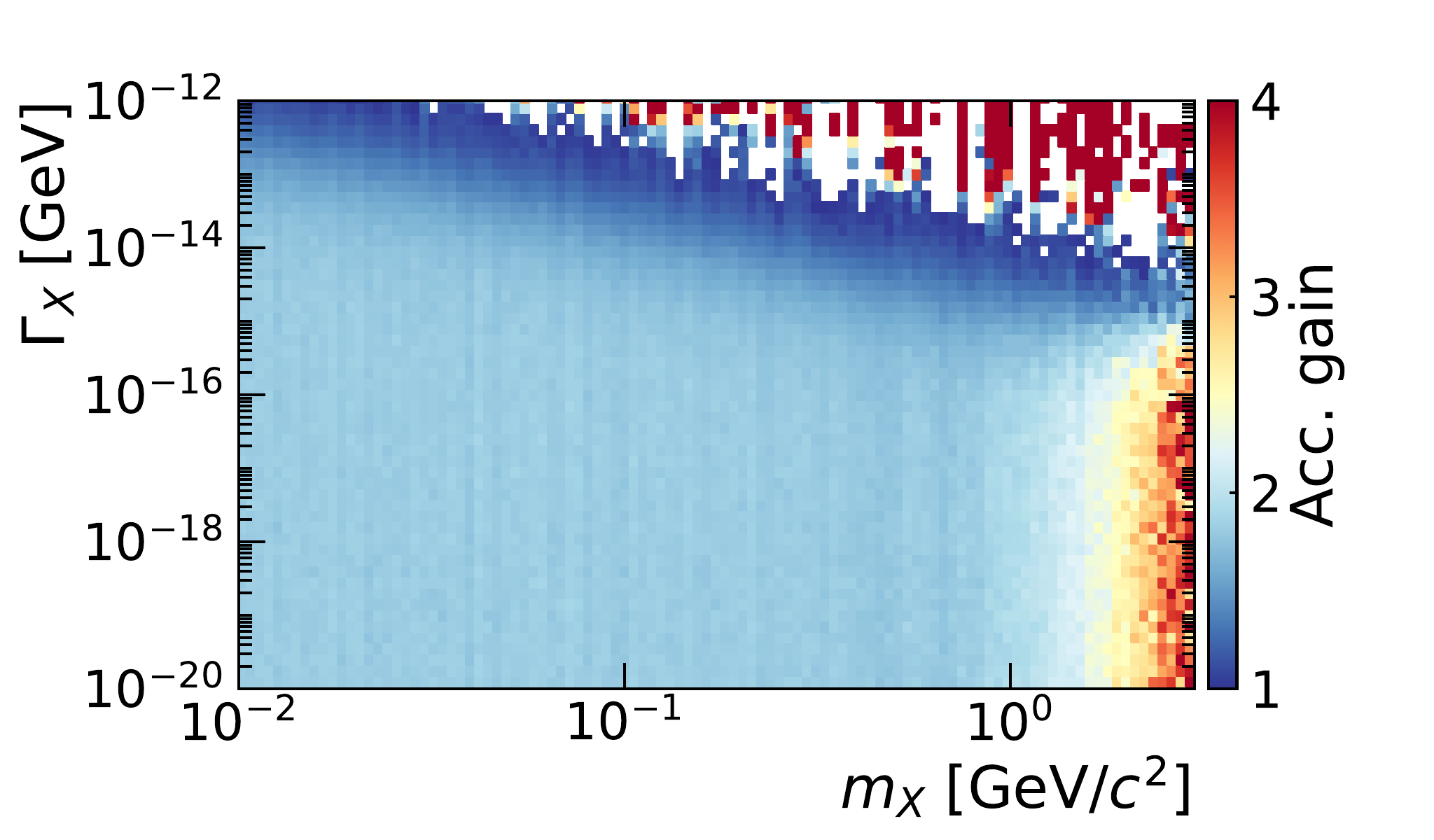}
    \caption{Ratio of the acceptance for setup \textbf{BDF 2a} to that for setup \textbf{BDF 0}, for a dark scalar decaying into $e^+ e^-$ produced by proton bremsstrahlung (left) and by \textit{B} meson decays (right).}
    \label{fig:bd2a}
\end{figure}

    \item 3a) The 17.69\,m long NA62 RICH detector is removed. The spectrometer is accordingly placed closer to the calorimeter system, allowing a longer decay volume than configuration 0 at the cost of losing the $\pi$, $K$ and $p$ separation, which is currently possible in certain momentum ranges at NA62. In this configuration, the STRAW stations are moved closer together, as in 2a) and the RICH is removed. This results in a 28.303\,m longer decay volume.
\end{itemize}

The impact of the geometry adjustment on the signal yield for direct and meson-mediated production processes can be seen in figures \ref{fig:bd3}, \ref{fig:bd1a} and \ref{fig:bd2a}. In summary, for configuration \textbf{BDF 1a} (\textbf{BDF 2a}), a global factor of 1.5 (2) increase in the signal yield can be expected with respect to the minimal configuration \textbf{BDF 0}. In configuration \textbf{BDF 3a}, the expected increase can reach a factor of 4.5 for direct production processes and a factor of 3.5 for meson-mediated processes.

\section{Fixed decay volume configurations \textbf{BDF 1b, 2b, 3b}}
\label{app:geo_b}

\begin{figure}[b]
    \centering
\includegraphics[width=0.49\textwidth]{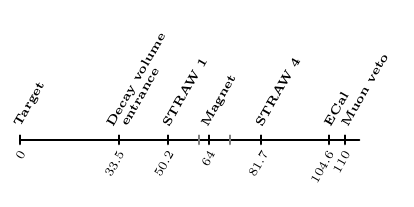}
\includegraphics[width=0.49\textwidth]{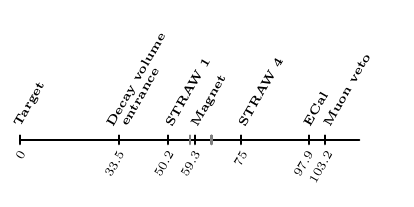}
\includegraphics[width=0.49\textwidth]{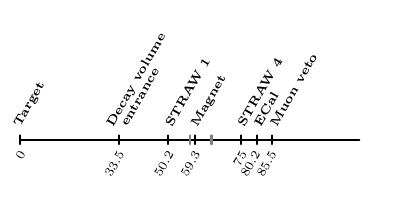}
    \caption{Location of the detectors and the decay volume along the $z$-axis with respect to the centre of the target in meters for \textbf{BDF 1b} (left), \textbf{BDF 2b} (right) and \textbf{BDF 3b} (bottom) configurations. The locations of the STRAW stations 2 and 3 are indicated in grey.}
    \label{fig:BDF123b}
\end{figure}

In these configurations, the position of the first STRAW station is considered fixed, while the modifications discussed in appendix \ref{app:geo_a} are applied, therefore keeping the same length of the decay volume while moving all detectors downstream of it closer to the target, as depicted in Fig.~\ref{fig:BDF123b}. This would effectively increase the signal yield by increasing the angular acceptance.

\begin{figure}[b!]
    \centering
\includegraphics[width=0.49\textwidth]{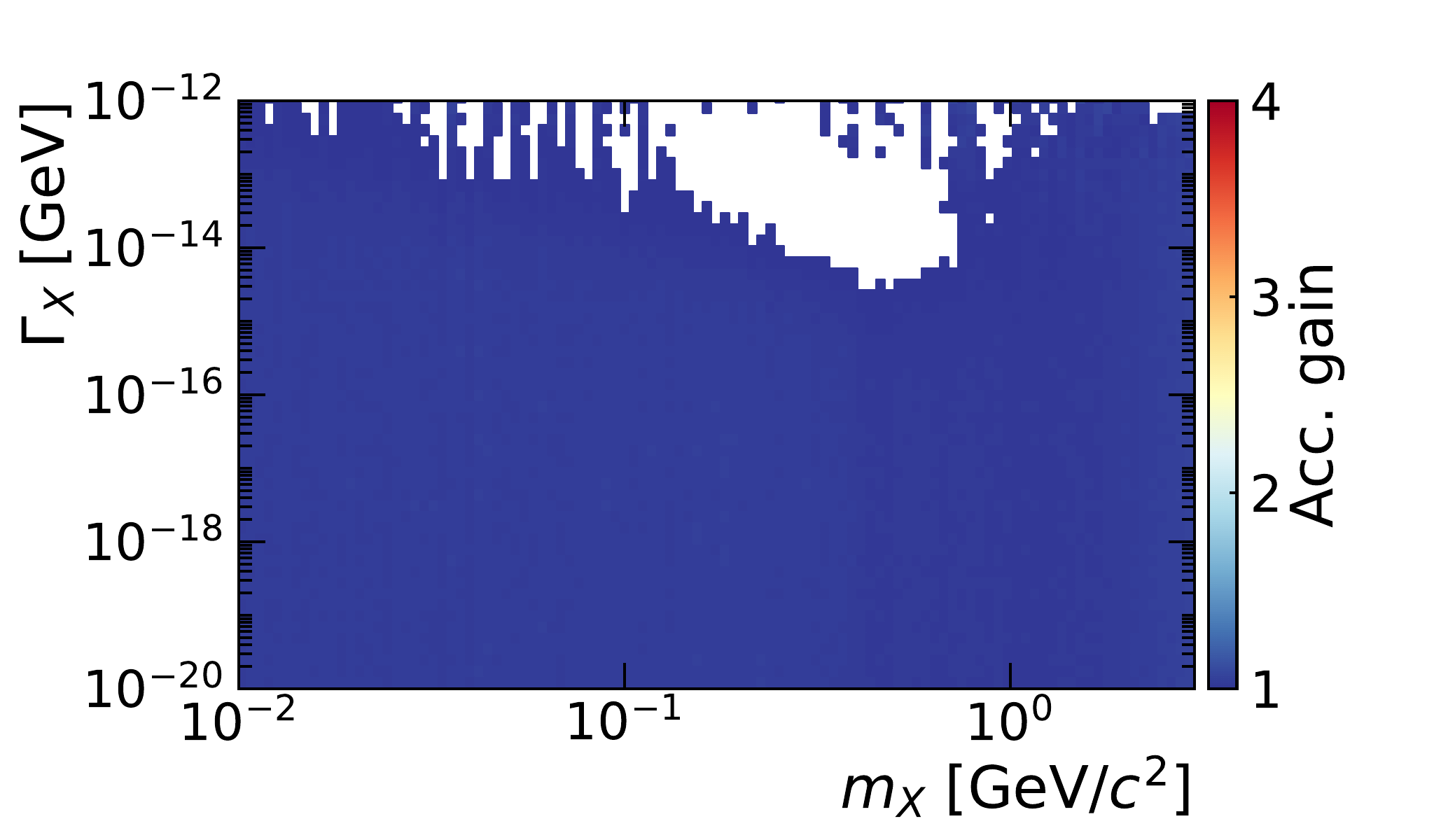}
\includegraphics[width=0.49\textwidth]{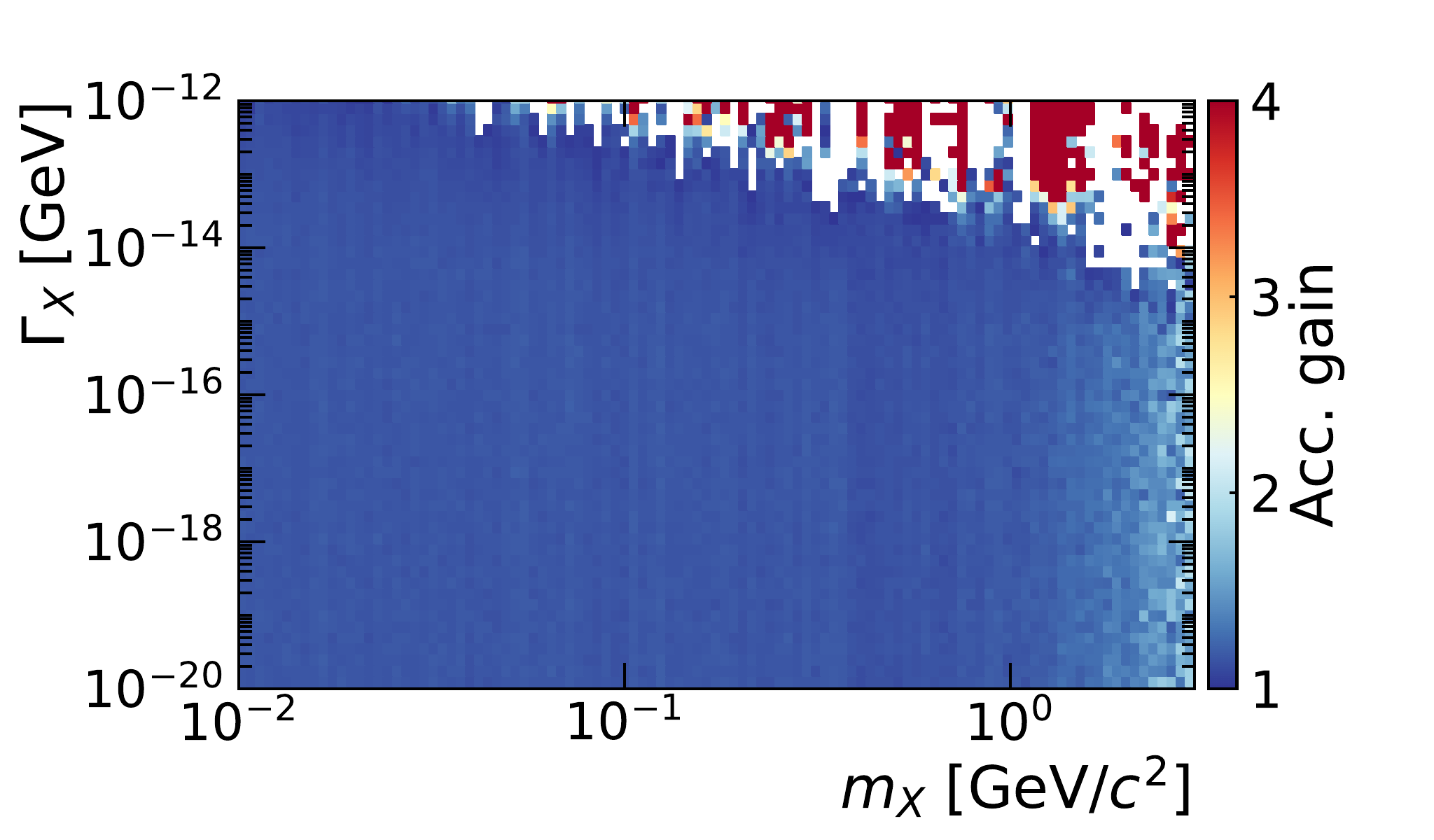}
\includegraphics[width=0.49\textwidth]{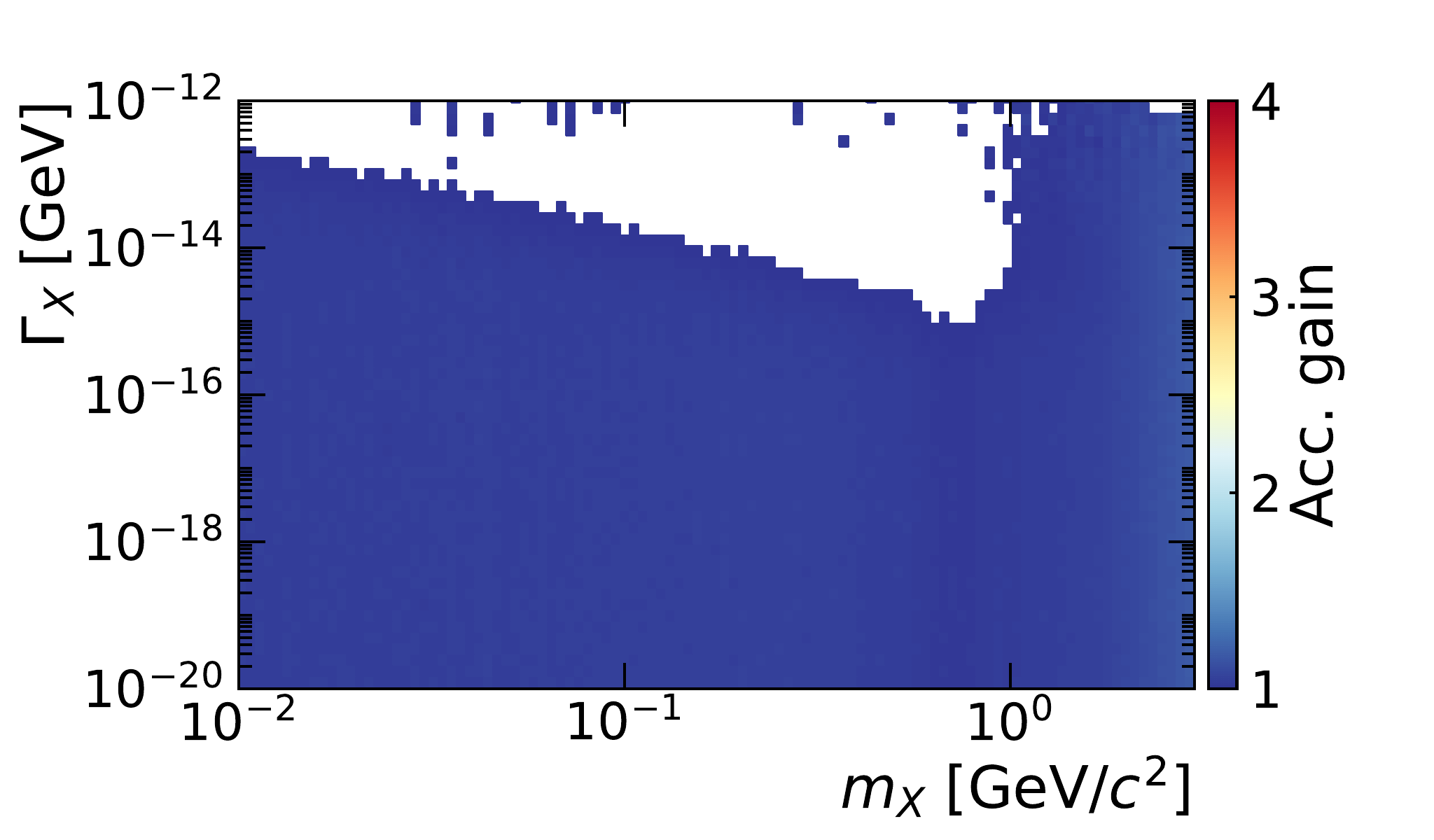}
\includegraphics[width=0.49\textwidth]{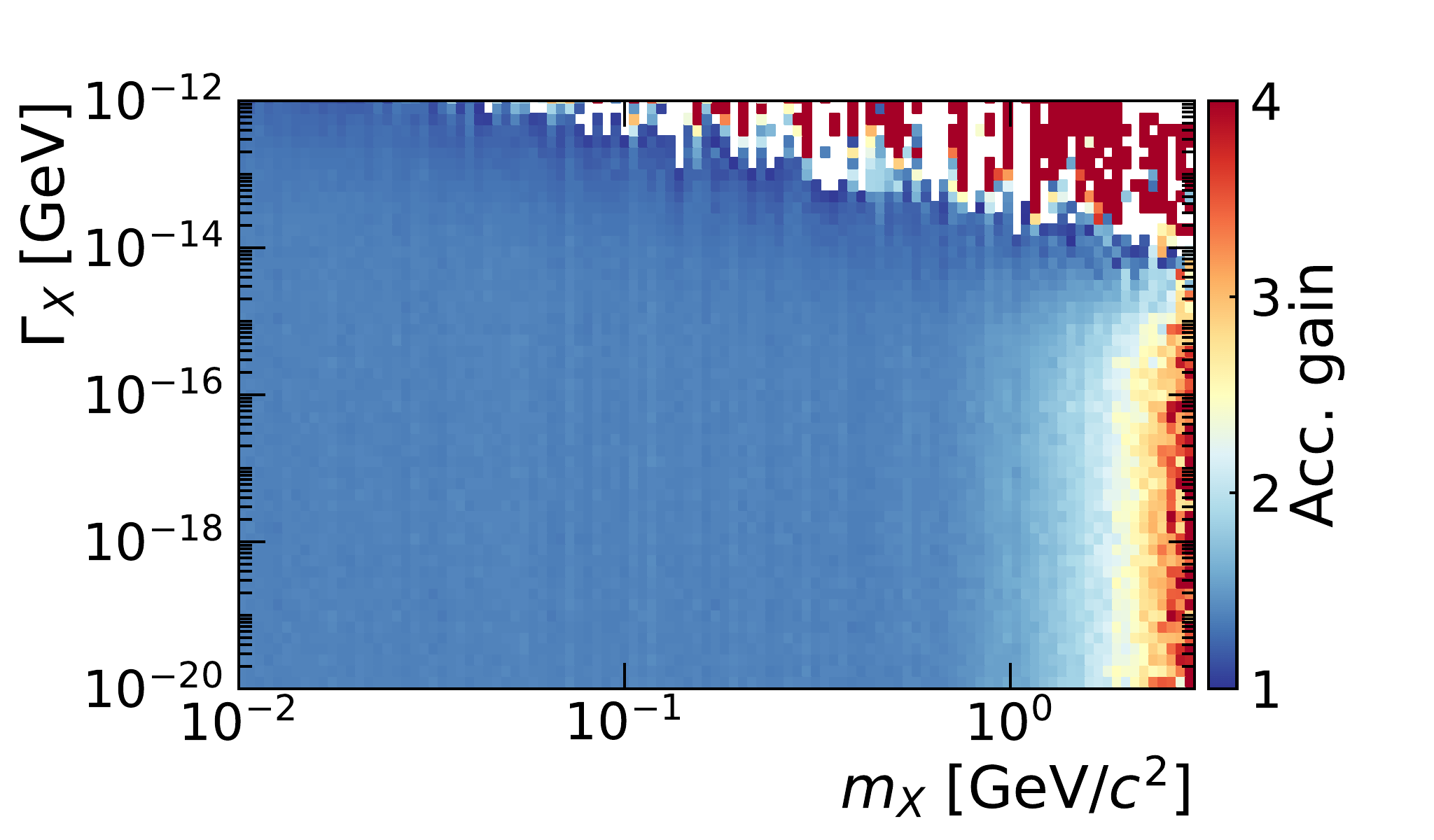}
\includegraphics[width=0.49\textwidth]{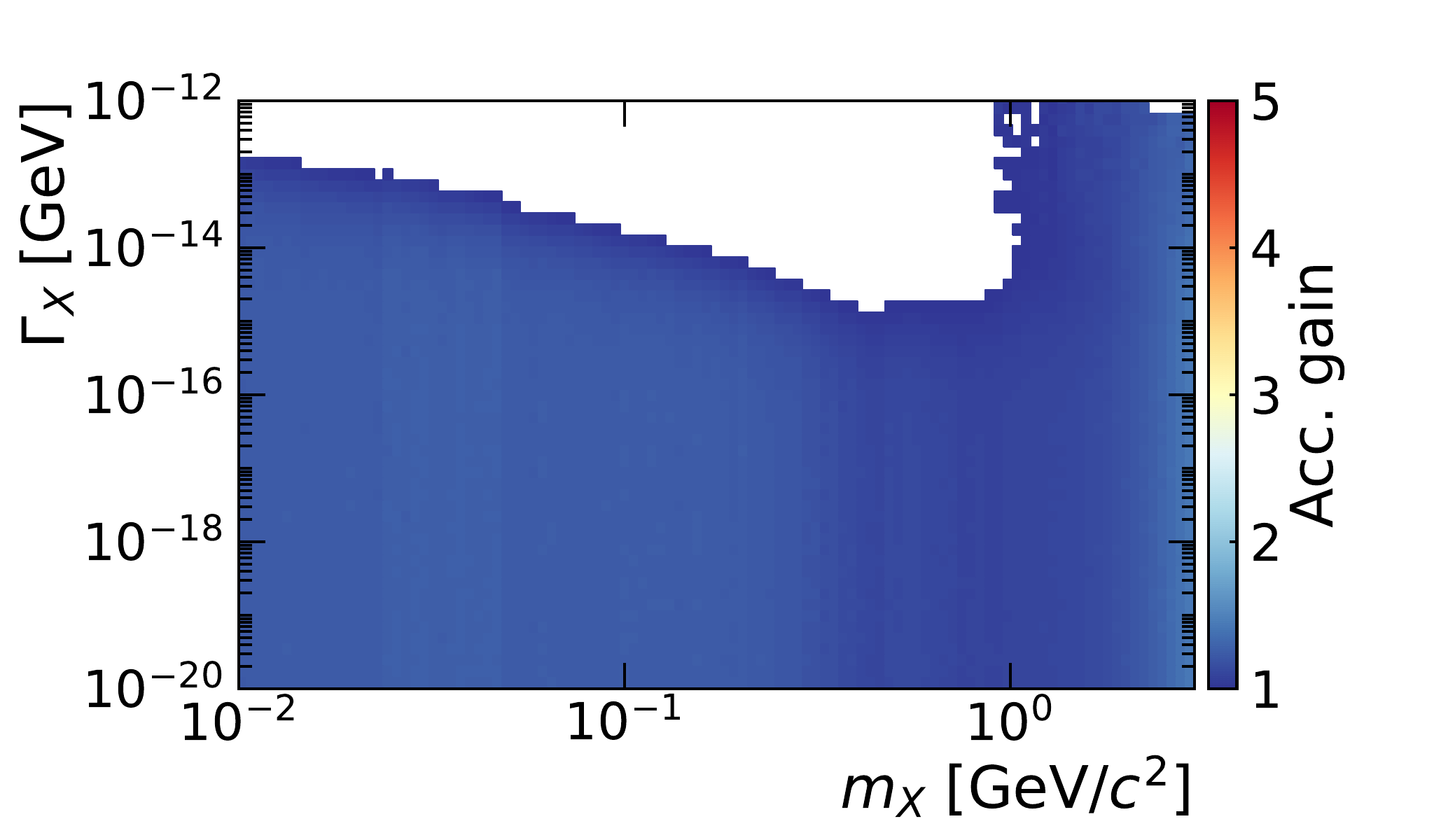}
\includegraphics[width=0.49\textwidth]{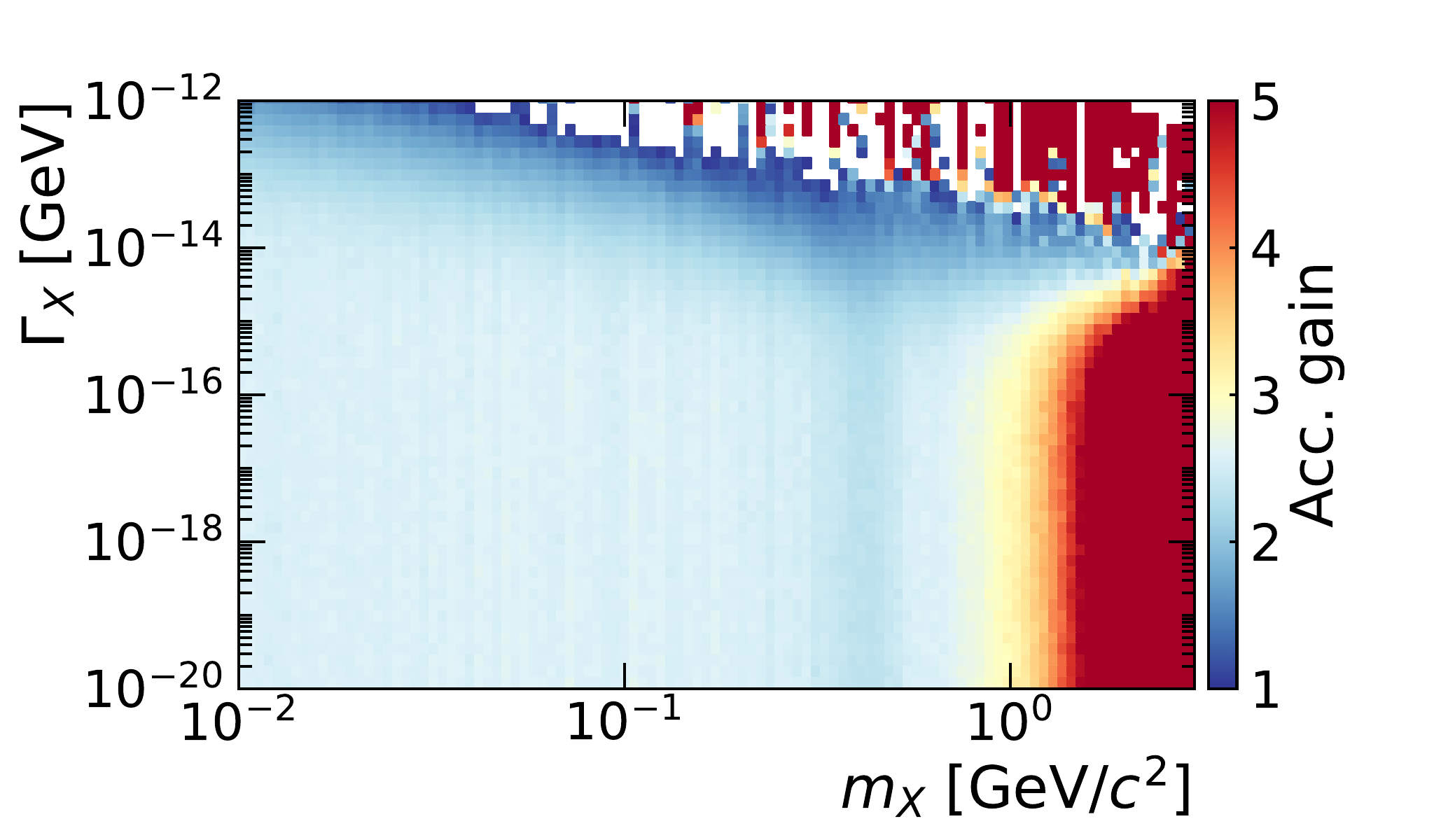}
    \caption{Ratio of the acceptance for setup \textbf{BDF 1b} (top), \textbf{BDF 2b} (centre) and \textbf{BDF 3b} (bottom) to that for setup \textbf{BDF 0} for a dark scalar decaying into $e^+ e^-$ produced by proton bremsstrahlung (left) and by \textit{B} meson decays (right).}
    \label{fig:bdb}
\end{figure}

In comparison with the \textbf{BDF 1a, 2a, 3a} configurations, the acceptance gain over configuration \textbf{BDF 0} is smaller, see Figures~\ref{fig:bdb}. While the meson-mediated productions can benefit from the larger angular acceptance, for direct production, the acceptance can even decrease when the calorimeter is too close to the decay volume because of the large boost and smaller separation between the final states at the calorimeter plane.

These configurations are also expected to be less favourable compared to the \textbf{BDF 1a, 2a, 3a} setups due to the complexity of the relocation of the ECal as well as the expected increase of the combinatorial background. Toy MC simulations are performed for these configurations anyway, for understanding the possible physics reach for different new physics scenarios.

\begin{table}[t]
    \centering
    \begin{tabular}{l|c|c|c}
    \hline
      scenario & $l_\text{decay volume}\,[\text{m}]$ & $z_\text{ECal}\,[\text{m}]$ & $\Omega_\text{ECal}\,[\text{msr}]$ \\ \hline\hline
      \rowcolor{lightgray}
      \textbf{BDF 0} & 16.7 & 108.5 & 0.4 \\
      \textbf{BDF 1a} & 20.6 & 108.5 & 0.4 \\
      \textbf{BDF 2a} & 27.3 & 108.5 & 0.4 \\
      \rowcolor{lightgray}
      \textbf{BDF 3a} & 45 & 108.5 & 0.4 \\
      \textbf{BDF 1b} & 16.7 & 104.6 & 0.43 \\
      \textbf{BDF 2b} & 16.7 & 97.9 & 0.49 \\
      \textbf{BDF 3b} & 16.7 & 80.2 & 0.73 \\
      \rowcolor{lightgray}
      \textbf{BDF 4} & 50 & 93.6 & 2.74 \\
      \hline
\end{tabular}
    
\caption{Overview of the main variables: the decay volume length, the distance between the target and the ECal and the solid angle covered by the ECal, which determine the detector acceptances for the configurations considered. The configurations which are displayed in the physics reach in section \ref{sec:reach} are highlighted.
    \label{tab:geo_overview}}
\end{table}

\section{Comparison of dark photon and dark scalar Bremsstrahlung}
\label{app:brems}

It has been shown in ref.~\cite{Foroughi-Abari:2021zbm} that for $E_\text{beam} = 120\,\mathrm{GeV}$ the differential probability for radiating the dark scalar is more sensitive to the off-axis angle than in the dark photon case. With higher centre of mass energies, demonstrated for 14\,TeV in ref.~\cite{Foroughi-Abari:2021zbm}, the difference becomes less pronounced but still apparent.

\begin{figure}[h]
    \centering
\includegraphics[width=0.49\textwidth]{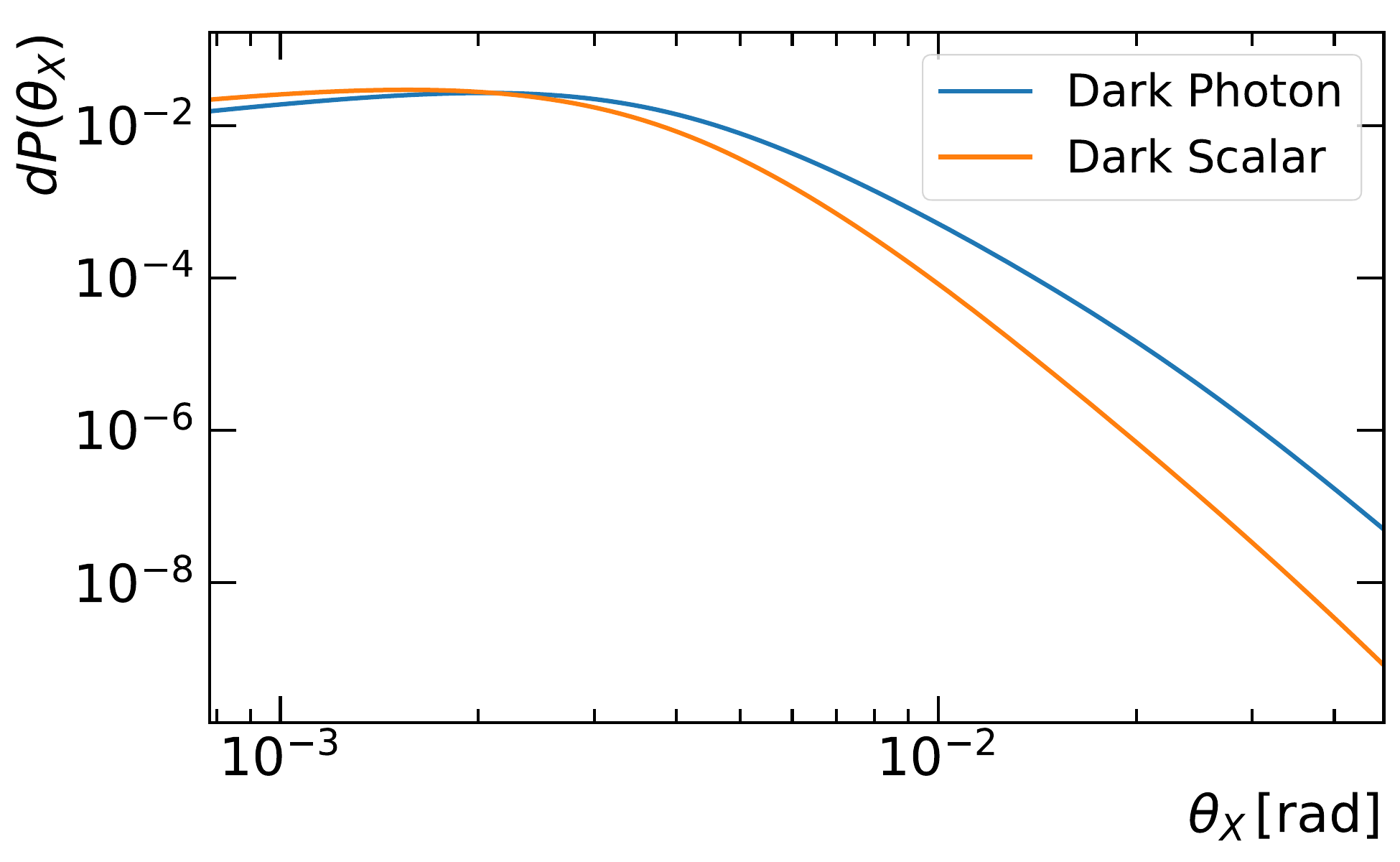}
    \caption{Bremsstrahlung production probability for a 400\,GeV proton beam on a fixed target as a function of the off-axis angle and integrated over dark photon and scalar masses and energies. The curve is normalized to 1.}
    \label{fig:brems}
\end{figure}

While we assume a 400\,GeV beam and use the more up-to-date ref.~\cite{Foroughi-Abari:2024xlj} for the dark photon, this feature persists, as shown in figure~\ref{fig:brems}. The effect is especially apparent in the sensitivity of the \textbf{BDF 0} on-axis and off-axis geometric configurations, where in the latter the centre of the calorimeter is about 12\,mrad off-axis with respect to the proton beam. The impact on the dark photon and dark scalar sensitivity can be compared in figures \ref{fig:BC1} and \ref{fig:BC45}.

\end{document}